\documentclass{aa}  

\usepackage{graphicx}
\usepackage{txfonts}
\usepackage[]{natbib}
\usepackage{color}

\begin{document}

   \title{Revealing the nature of central emission nebulae in the dwarf  galaxy NGC 185\thanks{Based on observations conducted with the 6m telescope of the Special Astrophysical Observatory of the Russian Academy of Sciences carried out with the financial support of the Ministry of Science and Higher Education of the Russian Federation, and on data collected with 2m RCC telescope at Rozhen National Astronomical Observatory.}}

   \author{M. M. Vu\v{c}eti\'{c}
          \inst{1}, D. Ili\'{c} \inst{1}, O. V. Egorov\inst{2,3}, A. Moiseev\inst{2,3,4},  D. Oni\'{c}\inst{1}, T. G. Pannuti\inst{5}, \\
          B. Arbutina\inst{1}, N. Petrov\inst{6}
                  \and
          D. Uro\v{s}evi\'{c}\inst{1,7}
          }

   \institute{Department of Astronomy, Faculty of Mathematics,
University of Belgrade, Studentski trg 16, 11000 Belgrade,
Serbia
 \and 
 Special Astrophysical Observatory, Russian Academy of Sciences, Nizhny Arkhyz 369167, Russia
 \and
Lomonosov Moscow State University, Sternberg Astronomical Institute, Universitetsky pr. 13, Moscow 119234, Russia  
\and
Space Research Institute, Russian Academy of Sciences, Profsoyuznaya ul. 84/32, Moscow 117997, Russia
\and
Department of Physics, Earth Science and Space Systems Engineering, Morehead State University, 235 Martindale Drive, Morehead, KY 40351, USA
\and
Institute of Astronomy and National Astronomical Observatory, Bulgarian Academy of Sciences, 72 Tsarigradsko Shosse Blvd, 1784 Sofia, Bulgaria
\and
Isaac Newton Institute of Chile, Yugoslavia Branch\\
\email{mandjelic@matf.bg.ac.rs}
             }

   \date{Received April xx, 2019; accepted xx xx, 2019}

  \abstract
  {} 
   {In this paper we present  new optical observations of the  galaxy NGC 185  intended to reveal the status of supernova remnants (SNRs)  in this dwarf  companion of the Andromeda galaxy. Previously, it was reported that this galaxy hosts one SNR.}
   { {Our deep photometric study with the 2m telescope at Rozhen National Astronomical Observatory using narrow-band H$\alpha$ and \hbox{[S\,{\sc ii}]} filters revealed complex structure of the interstellar medium in the center of the galaxy.} To confirm the classification and to study the kinematics of the detected nebulae, we carried out spectroscopic observations using the SCORPIO multi-mode spectrograph at the 6m telescope at the Special Astrophysical Observatory of the Russian Academy of Science, both in low- and high-resolution modes. We also searched the archival X-ray and radio data for counterparts of the candidate SNRs identified by our optical observations.}
   { {Our observations imply the presence of one more SNR, one possible \hbox{H\,{\sc ii}} region previously cataloged as part of an SNR, and the presence of an additional source of shock ionization in one low-brightness  PN.} We detected enhanced \hbox{[S\,{\sc ii}]}/H$\alpha$ and \hbox{[N\,{\sc ii}]}/H$\alpha$ line ratios, as well as relatively high  {(up to 90 km s$^{-1}$)} expansion velocities of the two observed nebulae, motivating their classification  as SNRs  {(with diameters of 45 pc and 50 pc)}, confirmed by both photometric and spectral observations.  {The estimated electron density of emission nebulae is 30 -- 200 cm$^{-3}$.} Archival {\it XMM-Newton} observations indicate the presence of an extended,  {low-brightness, soft} source in projection of one of the optical SNRs, {whereas the archival VLA radio image shows weak, unresolved emission in the center of NGC 185.}}
   {}

  
   \titlerunning{Revealing the nature of  emission nebulae in dwarf galaxy NGC 185}
    \authorrunning{Vu\v{c}eti\'{c} et al.}
   \maketitle
%


\section{Introduction} 
By studying emission nebulae in a galaxy we can reveal the global properties of a population of a certain type of nebula (\hbox{H\,{\sc ii}} regions, planetary nebulae, supernova remnants - SNRs), and also properties of the interstellar medium (ISM), such as abundances, temperature, and so on.  Extragalactic emission nebulae provide us with an opportunity to study objects that can be considered to be at almost the same distance, which allows us to investigate and compare them much more uniformly than the Galactic nebulae. 
Nearby dwarf galaxies are especially interesting due to our inability to observe these faint galaxies at greater distances. The importance of dwarf galaxies also lies in the fact that they are the most abundant type of galaxy and are crucial for our understanding of the formation and evolution of galaxies  {\citep{1998Mateo, 2003TolstoyVenn}}. 

 Three particular Local Group dwarf  galaxies, NGC 185, NGC 147, and NGC 205, are all satellites of  Andromeda, a large spiral  Galaxy. NGC 185 is at a distance of 617 kpc \citep{Ge2015, McConnachie2012} and although known to be a spheroidal galaxy, it is also known to contain gas \citep{YoungLo1997}. There is evidence that this galaxy has  a complex star formation history, which includes recent star formation activity. Such evidence is present in the form of numerous blue stars  \citep{Baade1951} and other Population I features, such as dust condensations, \hbox{H\,{\sc i}} gas, and an SNR 
\citep{MartinezDelgado1999, Martins2012, DeLooze2016}. 
It is known that star formation has occurred a few gigayears ago in the outer parts of  NGC 185, and a few million years ago in the central part \citep[see][]{MartinezDelgado1999, Geha2015}.
In particular, the minimum age of those blue  objects, now thought to be stellar associations and young clusters, can be estimated to  about 100 Myr, while the average star formation rate (SFR) for the last 1 Gyr in NGC 185 is estimated to be $6.6 \times  10^{-4}$ M$_{\sun}$ yr$^{-1}$. This is in  good agreement with mid- to far-infrared observations of this galaxy \citep{Marleau2010}, implying a mass of the gas of $3.0 \times 10^5$ M$_{\sun}$, which is consistent with the predicted mass feedback from dying stars formed in the last burst of star formation 100 Myr ago. Also, based on infrared spectral properties of NGC 185, \citet{Marleau2010} confirmed signatures of  star formation a few $10^8$ years ago.
\citet{Geha2015} concluded that the star formation in NGC 185 ceased sometime before, around  3 Gyr ago, based on their analysis of specific fields observed by the Hubble Space Telescope Advanced Camera for Surveys (at around 1.5 effective radii from the galaxy center, to avoid photometric crowding). These latter authors found that around 70\% of the stars in NGC 185 were formed more than 12.5 Gyr ago, with the majority of the remaining population forming between 8 to 10 Gyr ago. However, observations of the central region  of NGC 185 (confined to its inner 200 pc) show evidence for star formation as recent as 100 Myr ago, and a strong metallicity gradient, becoming more metal-poor with radial distance \citep{Crnojevic2014, Vargas2014}. \citet{Geha2015} suggested that the orbit of NGC 185  {around M31} has a sufficiently large perigalacticon, allowing it to preserve radial age and metallicity gradients and maintain a small central reservoir of recycled gas and dust.

Detection of SNRs in elliptical galaxies is relatively rare due to the fact that star formation ceased a long time ago. Therefore, NGC 185 provides a good opportunity to study evolution of this type of object in a low-density environment.  Optical extragalactic searches for SNRs predominantly use the fact that the optical spectra of SNRs have elevated \hbox{[S\,{\sc ii}]} $\lambda\lambda$6717,6731 \AA\AA \, to H$\alpha$ $\lambda$6563 \AA \, emission-line ratios, as compared to the spectra of normal {\hbox{H\,{\sc ii}}} regions. { This emission ratio is used to differentiate between shock-heated SNRs and
photoionized nebulae. Supernova remnants are expected to have \hbox{[S\,{\sc ii}]}/H$\alpha>0.4,$ but these are often considerably higher, while   {\hbox{H\,{\sc ii}}} regions have \hbox{[S\,{\sc ii}]}/H$\alpha<0.4$, but this is typically less than 0.2 \citep{MatonickFesen1997, BlairLong1997}}. Confirmation of SNRs is done by performing follow-up  spectroscopy of SNR candidates selected by narrow-band photometry, and with detection in other wavelengths. In addition, optical spectra allow the use of the so called BPT (after "Baldwin, Phillips \& Telervich"; \citealt{Baldwin1981}) emission line diagrams  to distinguish the ionization mechanism of nebular gas.  {It is generally accepted that the shock-heated objects (like active galactic nuclei (AGNs) and SNRs) and photoinized objects (like normal star-forming galaxies and {\hbox{H\,{\sc ii}}} regions) occupy different areas on these diagrams. In this work, we used  \hbox{[O\,{\sc iii}]}/H$\beta$ versus \hbox{[N\,{\sc ii}]}/H$\alpha$ and \hbox{[O\,{\sc iii}]}/H$\beta$ versus \hbox{[S\,{\sc ii}]}/H$\alpha$ BPT diagrams.}

Extragalactic optical searches for SNRs using emission-line ratio criterion  have so far detected more than 1200 SNRs  in 25 nearby galaxies, up to a distance of 10 Mpc \citep{Vucetic2015}. Nevertheless, only a small number of galaxies have been thoroughly surveyed for SNRs in more than one electromagnetic radiation range, and therefore it is hard to claim the
real status of SNRs in those galaxies. Detection of optical shock-heated emission-lines (e.g., \hbox{[S\,{\sc ii}]},  \hbox{[N\,{\sc ii}]},  \hbox{[O\,{\sc ii}]},  \hbox{[O\,{\sc i}]}), simultaneously with nonthermal radio spectral indices ($\nu^{-\alpha}$, $\alpha \sim 0.5$, with considerable dispersion around this value) and soft X-rays are suggestive of an SNR origin of the source (see e.g., \citealt{Filipovic1998, Bozzetto2017, Long2017}).

In this work, we aim to reveal the status of emission-line nebulae in NGC 185, with special attention devoted to  SNRs. 
For this purpose, we performed new optical deep observations of NGC 185,  {which are described in detail in the following sections.}  
Also, we searched archival, publicly available X-ray  and radio data in order to check for possible counterparts to optically detected objects. 
In Section 2 we summarize all previous observations of nebular emission in NGC 185; in Section 3 the optical observations and data analysis are described,  {while in Section 4 we present available, previously unpublished archival X-ray and radio-data; in Section 5 we present and discuss our results, and Section 6 outlines the main conclusions.}

\section{Previous observations of an SNR candidate in NGC 185}
\citet{Gallagher1984} were the first to notice that some of the emission in NGC 185 could originate from shock-heated plasma. These authors took long-slit spectra oriented east--west across the central part of the galaxy using the Kitt Peak National Observatory (KPNO) 4m Mayall telescope. According to the emission-line spectra, they concluded that the nebula, lying 0.5 \arcmin  to the east of the optical center of the galaxy, has properties similar to an SNR, due to the high intensities of \hbox{[S\,{\sc ii}]} and \hbox{[N\,{\sc ii}]} lines, compared to the H$\alpha$ line. They also discussed the fact that NGC 185 does not host any massive stars, meaning that this object cannot be a stellar wind nebula. A year later, \citet{Dickel1985} conducted radio observation of this SNR at 20 cm using the radio interferometer Karl G. Jansky Very Large Array (VLA). No obvious radio source was found at the expected area of the remnant. 
\citet{YoungLo1997} observed \hbox{H\,{\sc i}} and CO gas components of the NGC 185 galaxy. Also, they took  narrowband H$\alpha$ images of this galaxy. At the position of the possible SNR, they noticed extended H$\alpha$ emission, with crescent-shaped morphology,   17\arcsec = 50 pc in diameter. Its H$\alpha$ flux was estimated to $30 \pm 3 \times 10^{-15}$ erg cm$^{-2}$s$^{-1}$. These latter authors suggested that this emission is approximately coincident with the peak of \hbox{H\,{\sc i}} column density in NGC 185, and a comparison of the H$\alpha$ image with the locations of \hbox{H\,{\sc i}} clumps in NGC 185 showed that  {some} clumps surround the H$\alpha$ emission. 
\citet{MartinezDelgado1999} performed new H$\alpha$ observations of the central region of NGC 185 under good seeing conditions ($\sim$0.7\arcsec), which confirmed that the central emission nebula is extended and has an arc-like morphology, suggesting that it might be a portion of a larger, old remnant. According to its diameter, and assuming that this SNR is in Sedov-Taylor phase ( {adiabatic phase of evolution of an SNR, when  energy losses by radiation are very small, and when SNR diameter correlates with its age as $D \sim t^{2/5}$; \citealt{Sedov1959, Taylor1950})}, \citet{MartinezDelgado1999} estimated the age of this SNR to be $10^5$ years. Also, based on the absence of an upper main sequence on the color magnitude diagram of the stellar population in NGC 185, as well as  the assumption that the SFR for the last 100 Myr is  {negligible}, they concluded that this SNR originates from a type Ia event.

\citet{Goncalves2012} obtained deep spectroscopic observations of the H$\alpha$-emitting population of NGC 185 galaxy using the  multi-object spectrograph at the Gemini North telescope. These latter authors detected an emission-line object with a FWHM = 0.7\arcsec=2 pc (two times larger than their seeing), claiming that this object is a central part of the SNR detected by \citet{Gallagher1984}. For this object (ID 9-SNR1), \citet{Goncalves2012} measured \hbox{[S\,{\sc ii}]}/H$\alpha$ = 0.53. Since they did not detect any \hbox{[O\,{\sc iii}]} emission in the spectrum, these latter authors concluded that the shock velocity of this SNR is less than 85 km s$^{-1}$ \citep{Dopita1984}. This agreed well with their assumption that this is an old, evolved SNR. They also detected a portion of faint arc-like structure (see Table 1 from \citealt{Goncalves2012}, entry 11-ISM), characterizing it as the outer part of the SNR. The spectrum of this object showed only Balmer lines, and they did not characterize this faint diffuse nebula in more detail. Their derived  characteristics of the known SNR located close to the centre of NGC 185 were in contradiction with previous studies  {\citep{Gallagher1984,MartinezDelgado1999}}, firstly in terms of obtained \hbox{[S\,{\sc ii}]}/H$\alpha$ ratio, and also diameter.  This motivated us to obtain additional spectra at several positions of faint shell-like SNR, and other H$\alpha$-emitting objects in NGC 185, to reveal their  nature.

 {NGC 185 galaxy was undetected in the historical Very Large Array (VLA) survey of the nuclei of bright galaxies at 6 cm and 20 cm with the upper flux density limit of 1 and 5 mJy \citep{Heckman1980}. Later on, \citet{Dickel1985} searched for VLA counterparts of extragalactic optical SNRs at 20 cm. They set the upper limit to the integrated emission over 15\arcsec of the optical SNR candidate from \citet{Gallagher1984} to 1 mJy (rms noise $\sim$0.03 mJy/beam, beam size 4.6\arcsec$\times$3.6\arcsec). More recently, \citet{HoUlvestad2001} observed this galaxy again at 6 cm and 20 cm in 1999 using the VLA, and they set a higher  limiting flux density, integrated over the SNR diameter,  than previously set by \citet{Dickel1985}.}

 {NGC 185 was first observed in the X-rays by {\it ROSAT} \citep{Brandt1997}. Observation with total raw exposure of 21.0 ks pointed to the optical center of the galaxy but did not detect any X-ray sources coincident with the galaxy core. From the smoothed X-ray image no evidence for diffuse emission was seen. Recently, \citet{Ge2015} investigated the unresolved X-ray emission in NGC 185 using \textit{XMM-Newton} observations, in search of the X-ray properties of the old stellar population, while the discrete X-ray source population (including X-ray sources associated with SNRs) of this galaxy remained undetected.} 

An additional open question surrounding the investigation of NGC 185 galaxy is whether or not it hosts an AGN. \citet{Ho1997} measured the \hbox{[S\,{\sc ii}]}/H$\alpha$ ratio to be 1.5 in this galaxy, and they classified the NGC 185 nucleus as Seyfert 2, noting that shock ionization could also reproduce the observed line ratios. Since NGC 185 was not detected in radio surveys \citep{Heckman1980, HoUlvestad2001}, or X-ray observations \citep{Brandt1997}, it was suggested that the Seyfert-like line ratios are most probably produced by  stellar evolution processes \citep{Martins2012}. \citet{Martins2012} proposed that a mixture of SNRs and planetary nebulae (PNe) could be the source of the ionization for this galaxy. These latter authors showed that a composition of these two classes of objects mimic observed Seyfert-like line ratios, and it is now generally believed that this galaxy does not contain a genuine AGN. The work of these latter authors  supports the idea that more SNRs should be present in the galaxy, since their chemical evolution models suggest that around 80 SNe II and 15 SN Ia exploded in the last 100 Myr, with current SN rates being $8.8-9.4\times10^{-5}$ yr$^{-1}$ for SN II and $1.1-1.8\times10^{-5}$ yr$^{-1}$ for SN Ia. 

\subsection{Planetary nebulae in NGC 185}
The galaxy NGC 185  hosts at least eight PNe. The first five PNe were originally discovered by \citet{Ford1977}, and later confirmed by \citet{Corradi2005} and \citet{RicherMcCall2008}. \citet{Goncalves2012} detected three additional low-brightness PNe and the first symbiotic star in this galaxy. For the five brightest PNe, \citet{RicherMcCall2008} derived temperatures and elemental abundances, while  \citet{Goncalves2012} thoroughly discussed the   {physical gas condition}, as well as chemical abundances for the four brightest PNe, connecting them with the chemical evolution of NGC185.  \citet{Goncalves2012} for the first time derived electron densities for the three brightest PNe, which are somewhat higher than expected (ranging from 1800 to 16500 cm$^{-3}$), while, on the other side, they obtained $n_{\rm{e}}\sim 240$ cm$^{-3}$ for one low-brightness PN.  

\begin{figure}
  \centering
 \includegraphics[width=0.99\columnwidth]{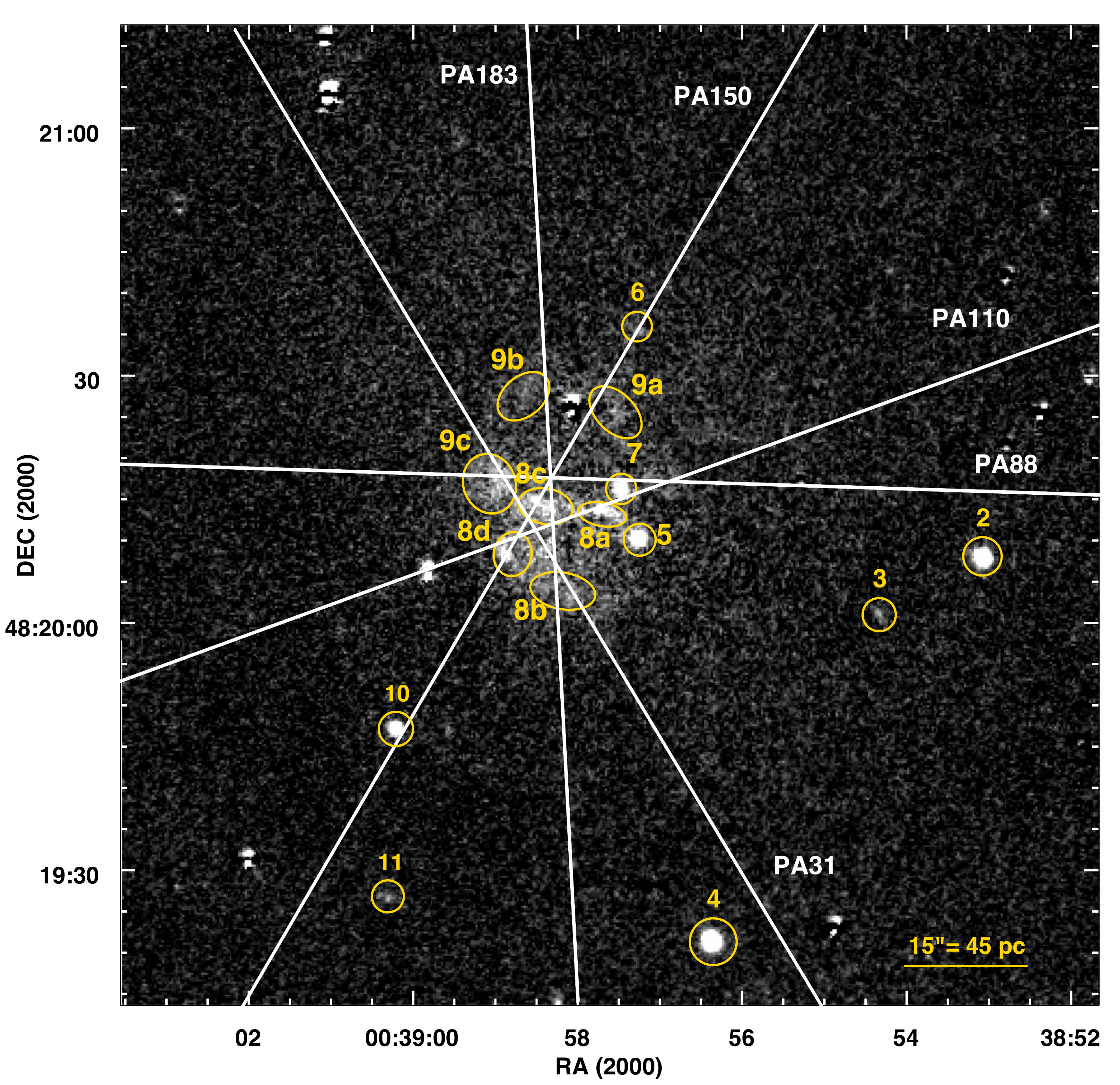}
\caption{{H$\alpha$ emission-line image (continuum subtracted) of the central $\sim 2 \arcmin \times 2 \arcmin$ of NGC 185 galaxy, with marked slit positions and emission-line objects. This figure is a zoom-in of the imaged portion of the galaxy,  intended to provide a better visualization of the central emission. Since object  {VIEM} 1 is located in the outer part of the galaxy, it is not visible here.}
              }
         \label{slits}
   \end{figure}

\section{Observations and data analysis}
\subsection{Optical photometry}
We observed the central part of NGC 185 galaxy using the 2m Ritchey Chretien-Coude (RCC) telescope at  NAO Rozhen. 
The telescope was equipped with VersArray: a 1300B CCD camera with a scale of 0.25\arcsec/px, giving a field of view of 5\arcmin45\arcsec $\times$ 5\arcmin35\arcsec. Observations were carried out  {during photometric nights} on November 2-3, 2015. 
The observations were performed through the narrowband \hbox{[S\,{\sc ii}]}, H$\alpha,$ and red continuum filters  {(see Table \ref{tab:filters} for filter characteristics, where $\lambda _o$ is the central wavelength of the filter, FWHM is the full-width half-maximum of the filter transparency curve, and  $\tau _\mathrm{max}$ is its maximum.)}.  We took sets of four images through each filter, with a total exposure time of 80 minutes per filter. Typical seeing was 1.25\arcsec– 2.25\arcsec. For the absolute flux calibration, we took images of spectrophotometric standard star BD+28 4211 \citep{Oke1990}.

\begin{table}
 \caption{Characteristics of the narrow band filters}
\label{tab:filters}
\begin{tabular}{l c c c }
\hline
Filter & $\lambda _o$ [\AA]  & FWHM [\AA]  & $\tau _\mathrm{max}$ [\%] \\
\hline
Red continuum & 6416 & 26 & 58.0 \\
H$\alpha$ & 6572 & 32 & 86.7 \\
\hbox{[S\,{\sc ii}]} & 6719 & 33 & 83.3 \\
\hline
        \end{tabular}
\end{table}

Data reduction was done using standard procedures in IRIS software\footnote{Available from http://www.astrosurf.com/buil/}. An astrometric calibration of the images was performed using the U.S. Naval Observatory’s USNO-A2.0 astrometric catalog \citep{Monet1998}. In order to obtain pure line emission, H$\alpha$ and \hbox{[S\,{\sc ii}]} images were continuum subtracted (scaling each image prior to subtraction to have the same flux as foreground stars) and corrected for filter transmission. Also, H$\alpha$ flux was corrected for contamination of \hbox{[N\,{\sc ii}]} lines at $\lambda$6548 \AA\ and $\lambda$6583 \AA using emission line  ratios  {of each source from our} spectroscopic observations,  {or  from \citet{Goncalves2012}, for objects which we could not be observed spectroscopically, such as for PNe.} { Galactic extinction was removed assuming that A(H$\alpha$)=0.6A$_{\rm{B}}$ and A$_{\rm{B}}$=0.667 \citep{Schlafly2011}.} See \cite{Vucetic2013} for all details on the procedure of data reduction and flux calibration.

\subsection{Optical spectroscopy}

\begin{table}
 \caption{Log of spectral observational data}
\label{tab:spec_data}
\begin{tabular}{llllll}
\hline
        PA,$^{\circ}$   & Date of obs.    & $\mathrm{T_{exp}}$, s  & $\theta$, \arcsec   & $\Delta\lambda$, \AA & $\delta\lambda$, \AA  \\ 
                \hline
        88 & 2017 Jul 27 & 6000 & 1.5 & {3650--7740} & 14 \\
        150 & 2017 Jul 27 & 6000 & 1.8 & {3650--7740} & 14 \\
        31 & 2017 Sep 25 & 3600 & 1.1 & {6080--7000} & 3.6 \\
        110 & 2017 Sep 25 & 7200 & 1.0 & {6080--7000} & 3.6 \\ 
        183 & 2017 Sep 25 & 8400 & 1.5 & {6080--7000} & 3.6 \\
                \hline
        \end{tabular}
\end{table}

 \begin{figure*}[ht]
   \centering
  \includegraphics[width=15cm]{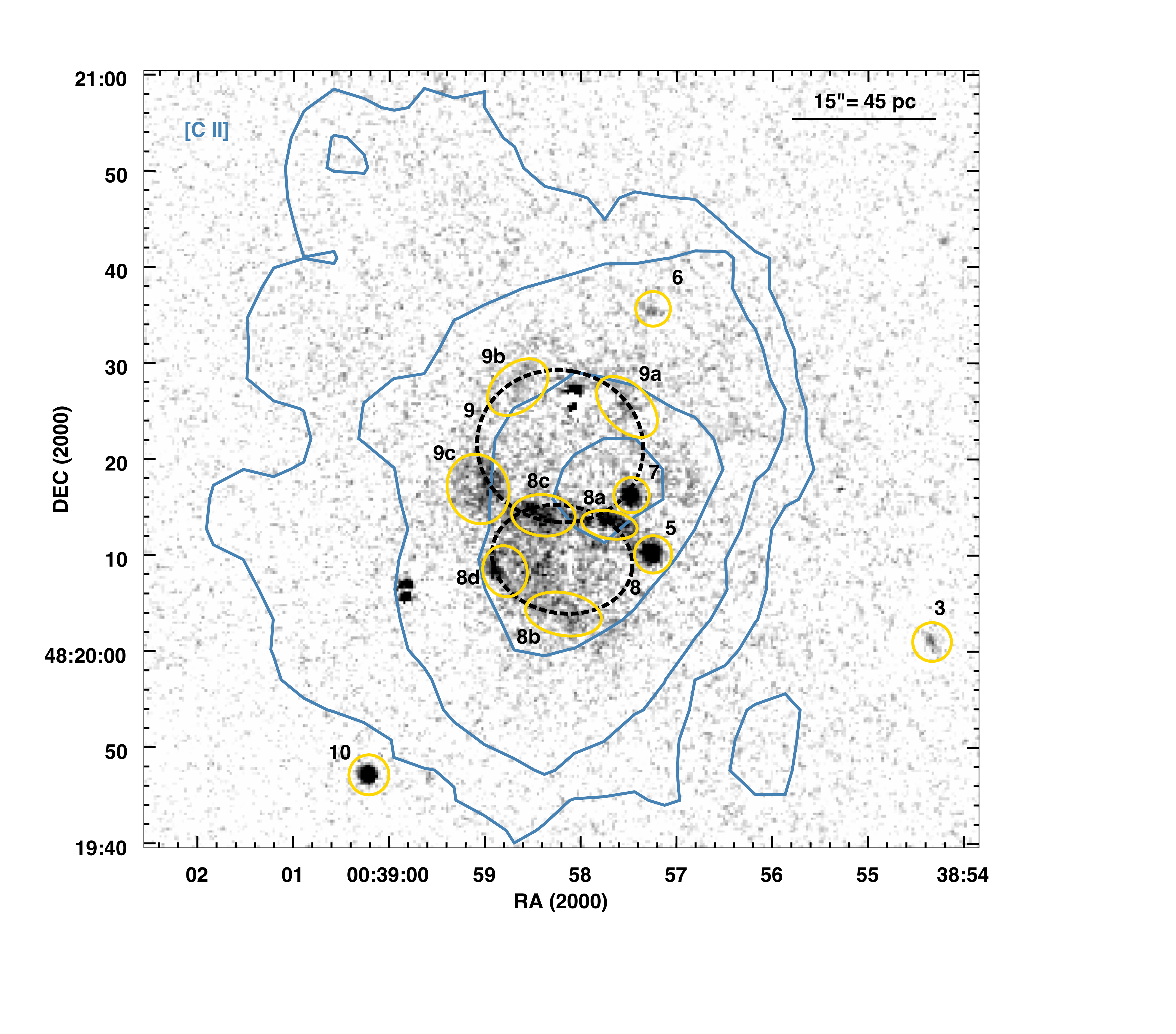}
   \caption{Zoom-in of the continuum-subtracted H$\alpha$ image of the center  ($\sim$1.5\arcmin$\times$1.5\arcmin)   of NGC 185 with \textit{Herschel}  PACS \hbox{[C\,{\sc ii}]} contours overlaid, indicating the location of dense gas. Objects are marked with full circles, and dashed circles follow the shell-like structure of objects  {VIEM} 8 and 9.}\label{n185-CII}
\end{figure*}

The long slit observations were carried out with the SCORPIO multi-mode spectrograph \citep{scorpio} at the prime focus of the SAO RAS 6m telescope. The  slit length  was 6 \arcmin and  its width was 1\arcsec. Two spectra at low resolution were obtained with the grism VPHG550G in the whole optical spectral range, while the VPHG1800R grism was used for observations with higher resolution in order to study the ionized gas kinematics in the spectral range including H$\alpha$, \hbox{[N\,{\sc ii}]}  and \hbox{[S\,{\sc ii}]} emission lines.  Table~\ref{tab:spec_data}  lists a log of the spectral observational data, where PA is  the position angle of the slit, $T_{\rm exp}$ is exposure time, $\theta$ is seeing, $\Delta\lambda$ is spectral range, and $\delta\lambda$ is the spectral resolution measured as the FWHM of air-glow emission lines. Figure~\ref{slits} shows the SCORPIO slit positions   overlaid on a H$\alpha$ image of the galaxy.

Data reduction was performed in a standard way using \textsc{idl}-based software written for reducing long-slit spectroscopic data obtained with SCORPIO. The main steps of the data reduction include bias subtraction,
geometric distortion and flat field corrections, linearization, and night-sky line subtraction. The reference spectrum of a He–Ne–Ar lamp obtained during the observations was used for linearization  {and wavelength calibration}. We observed the spectrophotometric standards BD+25d4655 (Feige 110 for PA=88$^\circ$) at a similar airmass immediately after the object and used them to calibrate the spectra to the absolute intensity scale.

To subtract the spectra of the stellar population we used a template spectrum constructed from the area outside the region of the observed emission lines for each slit position. This template was fitted to the observed spectrum at each pixel along the slit and the resulting model was subtracted. To measure the fluxes of emission lines we used our package  in \textsc{idl}  based on the \textsc{mpfit} \citep{mpfit} routine. Single-component Gaussian fitting was applied to measure the line flux distribution along the slits. To estimate the final uncertainties we quadratically added the errors propagated through all data-reduction steps to the uncertainties returned by \textsc{mpfit}. The reddening correction was applied to each spectrum before  estimating line fluxes. For that we derived the color excess $E(B-V)$ from the observed Balmer decrement ( {using only H$\alpha$/H$\beta$}) and then used the \cite{Cardelli1989} curve to perform reddening correction.

\begin{table*}
\caption{Optically detected emission-line objects in NGC 185 galaxy.}             
\label{tab:objects}      
\centering                          
\begin{tabular}{l c c c c c c c c}        
\hline\hline            
 & RA & Dec & H$\alpha$ flux\tablefootmark{a} & \hbox{[S\,{\sc ii}]} flux & $D$\tablefootmark{b} & cross- &\\  
ID & (h:m:s)& (d:m:s) & [$10^{-15}$ & [$10^{-15}$ & [pc] & ID\tablefootmark{c}& Comment\\
& J2000 & J2000 & erg cm$^{-2}$s$^{-1}$] & erg cm$^{-2}$s$^{-1}$] & \\
\hline                        
 {VIEM} 1 &       00:38:47:9      &       48:19:32        &        {$0.8\pm 0.1$}   &       -       &       <6      &       5-PN5   &       PN      \\
 {VIEM} 2 &       00:38:53.1      &       48:20:08        &        {$8.9\pm 0.1$}   &       -       &       <6      &       1-PN1   &       PN      \\
 {VIEM} 3 &       00:38:54.4      &       48:20:01        &        {$0.7\pm 0.1$}   &       -       &       <6      &       10-PN8  &       PN      \\
 {VIEM} 4 &       00:38:56.4      &       48:19:21        &        {$14.9\pm 0.1$}   &       -       &       <6      &       4-PN4   &       PN      \\
 {VIEM} 5 &       00:38:57.3 &    48:20:10        &        {$7.0\pm 0.1$}   &       -       &       <6      &       2-PN2   &       PN      \\
 {VIEM} 6 &       00:38:57.3      &       48:20:35        &        {$0.2\pm 0.1$}   &       -       &       6       &       8-PN7   &         {contribution of shocks} \\
 {VIEM} 7 &       00:38:57.5      &       48:20:16        &        {$2.0\pm 0.1$}   &       $1.0\pm 0.1$    &       <6\tablefootmark{d}     &       9-SNR1  &  {\hbox{H\,{\sc ii}} region}    \\
 {VIEM} 8 &       00:38:58.2      &       48:20:09        &       $28.7\pm5.0$    &       $35.8\pm5.0$    &       45      &               &       SNR     \\
 {VIEM} 9 &       00:38:59.0      &       48:20:17        &        {$6.0\pm5.6$}    &       $2.4\pm5.6$     &       50      &               & SNR     \\
 {VIEM} 10        &       00:39:00.2      &       48:19:47        &        {$3.3\pm 0.1$}   &       -       &       <6      &       3-PN3   &       PN      \\
 {VIEM} 11        &       00:39:00.3      &       48:19:27        &        {$0.6\pm 0.1$}   &       -       &       <6      &       13-StSy &       symbiotic
star    \\

\hline                                   
\end{tabular}
\tablefoot{
\tablefoottext{a}{Flux
has been corrected for \hbox{[N\,{\sc ii}]} emission, as described in the text, and for Galactic extinction, assuming that A(H$\alpha$)=0.6A$_{\rm{B}}$; A$_{\rm{B}}$=0.667 \citep{Schlafly2011}}
\tablefoottext{b}{1\arcsec=3 pc for NGC 185 distance of 617 kpc}
\tablefoottext{c}{IDs from \citet{Goncalves2012}} \tablefoottext{d}{\citet{Goncalves2012} give estimate of 2 pc.}
}
\end{table*}

\section{Archival data}

\subsection{Archival X-ray data}

NGC 185 was observed with the three European Photon Imaging Cameras (EPIC) --  {Metal Oxide Semi-conductor} cameras (MOS1), MOS2, and  {pn-CCD} (PN) aboard the \textit{XMM-Newton} Observatory with a start date of 27~December~2010 (PI: Li Zhiyuan; Obs.~ID 0652210101). The aim of these latter authors was to examine the unresolved X-ray emission of the old stellar population in NGC 185 \citep{Ge2015}, so their analysis excluded all point sources in the field of view; we re-analyzed their archival data trying to extract any signature of the SNR emission.

We downloaded the archival dataset from the \textit{XMM-Newton} Science Archive and analyzed it using standard tools in the HEASOFT Software Package and the Science Analysis Software (SAS) software package (Version 17.0.0). The SAS tools \verb"epchain" and \verb"emchain" were used to apply standard processing tools to the EPIC data, while the tools \verb"pn-filter" and \verb"mos-filter" were used to filter the datasets for background flaring activity. After processing, the effective exposure times of the MOS1, MOS2, and PN cameras were 93 ks, 98 ks, and 81 ks, respectively. Combined EPIC (MOS1+MOS2+PN) images of NGC 185,  exposure-corrected and adaptively smoothed, were created using the Extended Source Analysis Software package \citep{SK11}. 

Also, we applied a simple analysis to get an impression of the X-ray spectrum of one source. Inspection of the extracted spectra revealed that all of the EPIC, MOS1, and MOS2 spectra lacked sufficient a signal-to-noise ratio for a detailed spectral analysis: therefore, we only considered PN spectra for our simple analysis.  Spectra were extracted using the SAS tool \verb"evselect". The background region was determined using the \verb"ebkgreg" tool, excluding all detected point sources that may lay inside that area. As W-statistic \citep[see e.g.,][]{ducci14} requires at least one count per spectral bin, we binned the data appropriately. The main focus was on the energy range between 0.4 keV and 3 keV, due to the soft nature of the X-ray source  and high soft proton contamination at high energies. 
Using \verb"nH" function in XSPEC, we adapted the hydrogen column density value of $N_{\rm H, Gal}=1.11\times10^{21}\ \mathrm{cm^{-2}}$ for photo-electric foreground absorption by the Galaxy. Actually we used the product of two Tuebingen-Boulder (\verb"TBABS") ISM X-ray absorption models \citep{tbabs00}, one representing the fixed Galactic absorption ($N_{\rm H, Gal}$) and a second taking into account possible intrinsic absorption in NGC 185 ($N_{\rm H, int}$). Only simple emission models were used. In all cases, abundance in NGC 185 was set to 0.6 of the solar abundance value. The first model that we used is the Astrophysical Plasma Emission Code \citep[\texttt{APEC},][]{apec01} for the optically thin plasmas in collisional ionization equilibrium (CIE).  We also tried to fit the PN spectrum with a single temperature and single ionization age nonequilibrium ionization  (\verb"NEI") collisional plasma model .

\begin{table*}
\caption{Observed line fluxes in $\mathrm{10^{-17}\, erg\, s^{-1}\, cm^{-2}\, arcsec^{-2}}$ for  {selected} emission-line objects measured from their integrated spectra within the denoted position ( {X-axis} marked with $R$ on Figs. \ref{pa31} -- \ref{pa183}) along corresponding slit. Reddening $E(B-V)$ is derived from the Balmer decrement (for low-resolution spectra), and electron density $n_\mathrm{e}$ is derived from \hbox{[S\,{\sc ii}]}6717/6731 line ratios  (for high-resolution spectra).}\label{table3}  
\centering
\footnotesize
\begin{tabular}{@{\extracolsep{-1.5mm}}ccccccc@{}}
Region & {VIEM} 6 (PA=150) & {VIEM} 7 (PA=88) & {VIEM} 7 (PA=110) & {VIEM} 8a (PA=110) & {VIEM} 8b (PA=31) & {VIEM} 8b (PA=183) \\
\hline
Position, \arcsec & $-14.3...-11.4$ & $-10.4...-6.1$ & $-1.1...1.7$ & $2.8...7.1$ & $-21.8...-18.2$ & $6.1...8.2$ \\
H$\beta$ & $<2.7$ & $3.2\pm0.4$ & $-$ & $-$ & $-$ & $-$ \\
\hbox{[O\,{\sc iii}]}4959 & $2.3\pm0.4$ & $0.5\pm0.1$ & $-$ & $-$ & $-$ & $-$ \\
\hbox{[O\,{\sc iii}]}5007 & $6.9\pm0.9$ & $1.4\pm0.3$ & $-$ & $-$ & $-$ & $-$ \\
\hbox{[N\,{\sc ii}]}6548 & $2.5\pm0.6$ & $2.1\pm0.1$ & $1.5\pm0.2$ & $1.9\pm0.2$ & $1.2\pm0.3$ & $1.3\pm0.3$ \\
H$\alpha$ & $10.4\pm1.1$ & $13.1\pm0.3$ & $16.3\pm0.3$ & $12.0\pm0.3$ & $4.9\pm0.4$ & $5.7\pm0.5$ \\
\hbox{[N\,{\sc ii}]}6584 & $7.8\pm1.1$ & $6.3\pm0.3$ & $4.5\pm0.3$ & $5.7\pm0.3$ & $3.7\pm0.4$ & $3.9\pm0.5$ \\
\hbox{[S\,{\sc ii}]}6717 & $2.1\pm0.7$ & $4.7\pm0.3$ & $4.6\pm0.2$ & $8.6\pm0.2$ & $4.9\pm0.2$ & $6.1\pm0.3$ \\
\hbox{[S\,{\sc ii}]}6731 & $3.3\pm1.4$ & $2.8\pm0.3$ & $3.0\pm0.2$ & $5.5\pm0.2$ & $3.7\pm0.3$ & $4.1\pm0.3$ \\
$E(B-V)$ & $0.24\pm0.31$ & $0.32\pm0.12$ & $-$ & $-$ & $-$ & $-$ \\
$n_{\rm e}$, cm$^{-3}$ & $-$ & $-$ & $<200$ & $<200$ & $100\pm110$ & $-$ \\
\hline
\hline
Region & {VIEM} 8c (PA=31) & {VIEM} 8c (PA=110) & {VIEM} 8c (PA=150) & {VIEM} 8c (PA=183) & {VIEM} 8d (PA=110) & {VIEM} 8d (PA=150) \\
\hline
Position, \arcsec & $-10.0...-6.4$ & $8.2...11.0$ & $10.4...14.6$ & $-4.3...-0.7$ & $14.9...18.5$ & $17.5...20.3$ \\
H$\beta$ & $-$ & $-$ & $7.9\pm0.5$ & $-$ & $-$ & $8.6\pm1.1$ \\
\hbox{[O\,{\sc iii}]}4959 & $-$ & $-$ & $12.5\pm0.6$ & $-$ & $-$ & $12.8\pm0.7$ \\
\hbox{[O\,{\sc iii}]}5007 & $-$ & $-$ & $37.5\pm1.4$ & $-$ & $-$ & $38.5\pm1.9$ \\
\hbox{[N\,{\sc ii}]}6548 & $1.5\pm0.3$ & $1.5\pm0.2$ & $9.1\pm0.5$ & $1.6\pm0.2$ & $1.2\pm0.2$ & $8.3\pm0.6$ \\
H$\alpha$ & $7.0\pm0.5$ & $10.1\pm0.3$ & $45.1\pm1.0$ & $8.2\pm0.5$ & $8.1\pm0.2$ & $38.3\pm1.8$ \\
\hbox{[N\,{\sc ii}]}6584 & $4.7\pm0.5$ & $4.6\pm0.3$ & $28.0\pm0.9$ & $4.9\pm0.5$ & $3.7\pm0.2$ & $25.4\pm1.5$ \\
\hbox{[S\,{\sc ii}]}6717 & $5.0\pm0.3$ & $6.7\pm0.3$ & $38.6\pm1.4$ & $6.7\pm0.2$ & $6.0\pm0.2$ & $37.4\pm1.5$ \\
\hbox{[S\,{\sc ii}]}6731 & $3.8\pm0.3$ & $4.5\pm0.3$ & $25.9\pm1.4$ & $4.8\pm0.2$ & $4.3\pm0.2$ & $27.7\pm1.7$ \\
$E(B-V)$ & $-$ & $-$ & $0.60\pm0.06$ & $-$ & $-$ & $0.38\pm0.12$ \\
$n_{\rm e}$, cm$^{-3}$ & $105\pm115$ & $<200$ & $-$ & $40\pm70$ & $30\pm75$ & $-$ \\
\hline
\hline
Region & {VIEM} 9a (PA=150) & {VIEM} 9b (PA=183) & {VIEM} 9c (PA=31) & {VIEM} 9c (PA=88) & {VIEM} 10 (PA=150) \\
\hline
Position, \arcsec & $-5.4...-2.5$ & $-20.0...-15.0$ & $-5.7...0.0$ & $3.9...8.9$ & $42.8...45.7$ \\
H$\beta$ & $2.7\pm1.1$ & $-$ & $-$ & $2.4\pm0.4$ & $13.6\pm0.8$ \\
\hbox{[O\,{\sc iii}]}4959 & $1.2\pm0.5$ & $-$ & $-$ & $5.2\pm0.2$ & $76.0\pm1.6$ \\
\hbox{[O\,{\sc iii}]}5007 & $3.5\pm1.1$ & $-$ & $-$ & $15.6\pm0.4$ & $227.9\pm4.7$ \\
\hbox{[N\,{\sc ii}]}6548 & $2.2\pm0.6$ & $0.4\pm0.2$ & $1.2\pm0.2$ & $1.4\pm0.1$ & $10.6\pm0.4$ \\
H$\alpha$ & $10.1\pm1.5$ & $2.3\pm0.2$ & $8.2\pm0.4$ & $10.4\pm0.3$ & $65.8\pm1.3$ \\
\hbox{[N\,{\sc ii}]}6584 & $6.8\pm1.2$ & $1.4\pm0.2$ & $3.5\pm0.3$ & $4.3\pm0.3$ & $32.4\pm0.9$ \\
\hbox{[S\,{\sc ii}]}6717 & $7.8\pm1.4$ & $1.3\pm0.2$ & $3.5\pm0.2$ & $3.2\pm0.3$ & $1.1\pm0.5$ \\
\hbox{[S\,{\sc ii}]}6731 & $5.9\pm1.7$ & $0.9\pm0.2$ & $2.3\pm0.2$ & $2.5\pm0.3$ & $3.8\pm2.2$ \\
$E(B-V)$ & $0.24\pm0.36$ & $-$ & $-$ & $0.36\pm0.13$ & $0.45\pm0.05$ \\
$n_{\rm e}$, cm$^{-3}$ & $-$ & $<200$ & $<200$ & $-$ & $-$ \\
\hline
\end{tabular}
\end{table*}

\subsection{Archival radio data}
 {We searched the archival radio data for the counterpart radio emission in order to support our findings concerning SNR candidates. The VLA archive lists several more recent unpublished observations of NGC 185 which were part of other surveys. These were done in different VLA configurations, and therefore with different spatial resolution and observed frequency. The most suitable observing run for our science case was done on July 20, 2001 (Project id. AY123), with VLA in C configuration at 1.4 GHz, with a beam size of 14.4\arcsec and an rms of 0.074 mJy/beam. Since the VLA archive hosts processed images from these observations, we performed only a rudimentary analysis in DS9 to obtain surface brightness estimates of the detected objects.}

\section{Results and discussion}

\subsection{Optical photometry  {results}}
In order to reveal more details about the SNR candidate in the galaxy NGC 185, we performed narrowband photometry through H$\alpha$ and \hbox{[S\,{\sc ii}]} filters. 
The photometric observations detected 11 objects emitting in H$\alpha$ and/or \hbox{[S\,{\sc ii}]} lines (Fig. \ref{slits}).
 {The objects were named VIEM, according to the initials of the first four authors\footnote{Following the IAU naming policy https://www.iau.org/public/themes/naming/.}, and numbered according to increasing right ascension:  {VIEM} 1--11.} We note that object  {VIEM} 1 (see Fig. 1 in \citealt{Vucetic2016}, object 1)  is located  {in the outskirts}, and is thus not shown in Fig. \ref{slits}.
In Table \ref{tab:objects} we list the coordinates of the objects as well as their H$\alpha$ and \hbox{[S\,{\sc ii}]} line fluxes and diameters. 

Among these detected objects are the already known optical SNR candidate (VIEM 8) and a possible new one (VIEM 9), which were revealed according to their high \hbox{[S\,{\sc ii}]}/H$\alpha$ flux ratio ($>0.4$). Figure \ref{n185-CII} shows a zoom-in of the continuum-subtracted H$\alpha$ image of the central part ($\sim$1.5\arcmin$\times$1.5\arcmin) of NGC 185 where the complex emission of  {optical SNRs}  {VIEM} 8 and  {VIEM} 9 is seen  {(traced with dashed circles, Fig. \ref{n185-CII})}. Both objects show partial shell-like structure with sub-components of enhanced emission, enumerated with 8a -- 8d, and 9a -- 9c. These filaments were confirmed and studied in more detail with spectral observations of high spatial and spectral resolution (see Section 5.2).  {Also, we propose that this galaxy hosts a compact \hbox{H\,{\sc ii}} region (VIEM 7) that is smaller in  size than our photometric resolution, and whose classification is  based on its position on BPT diagrams and absence of gas motion (Section 5.5).}  In addition, in Fig. \ref{n185-CII} we overlaid the \textit{Herschel} PACS \hbox{[C\,{\sc ii}]} contours from \cite{DeLooze2016} in order to show the distribution of dense gas,  {whose peak coincides with the position of objects  {VIEM} 7 and  {VIEM} 8a}.

 {In addition,} we detected seven previously known PNe. For one of them, designated PN-7 by \citet{Goncalves2012}, which is our object  {VIEM} 6, BPT diagrams  suggest possible shock ionization (see more details in Section 5.5). In addition, very low-brightness emission is seen both in H$\alpha$ and \hbox{[S\,{\sc ii}]}, at the position of a previously detected symbiotic star \citep{Goncalves2012}.
 
\subsection{Optical spectroscopy  {results}}

\begin{figure}
   \centering
   \includegraphics[width=0.8\columnwidth]{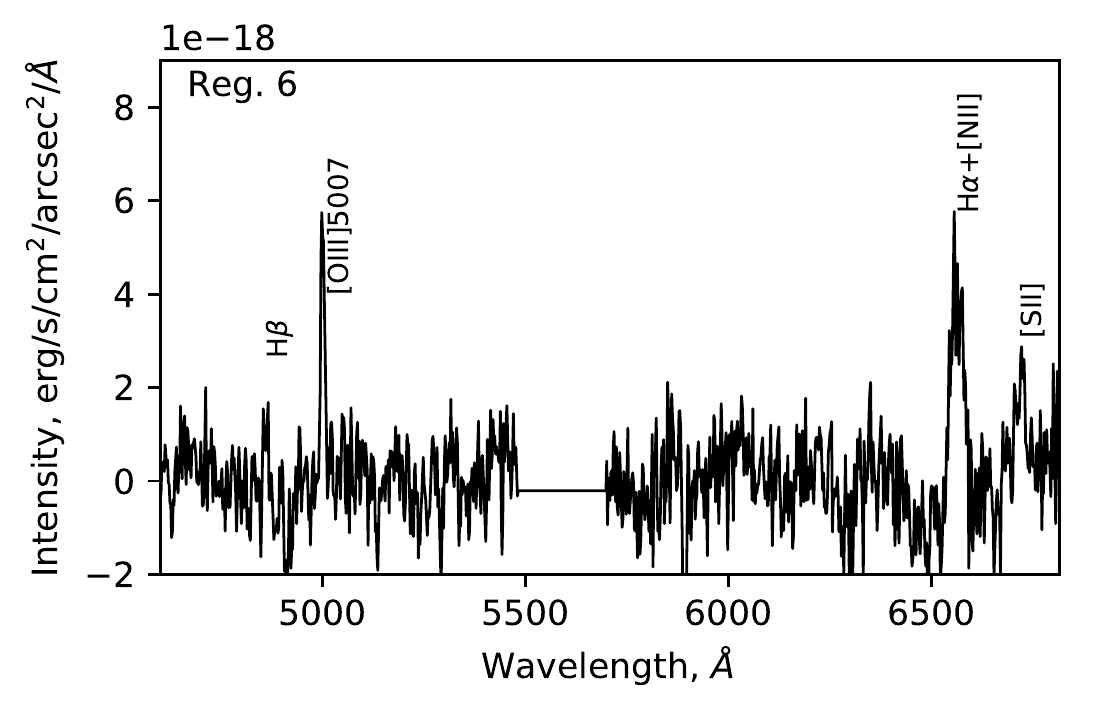}
    \includegraphics[width=0.8\columnwidth]{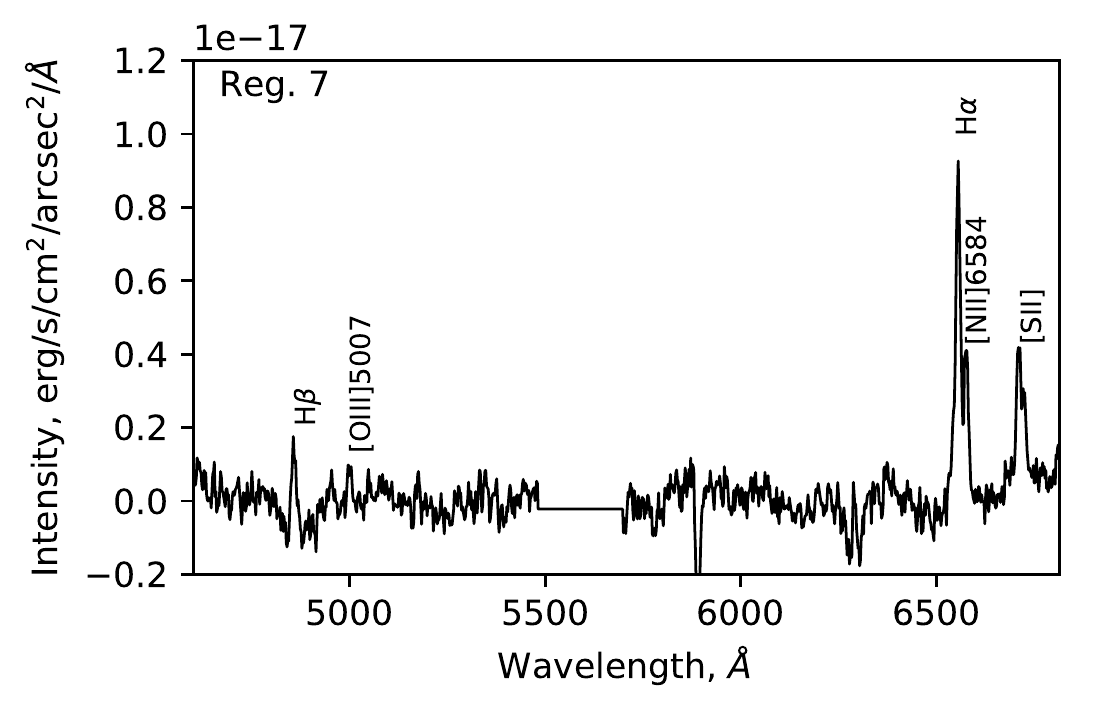}
    \includegraphics[width=0.8\columnwidth]{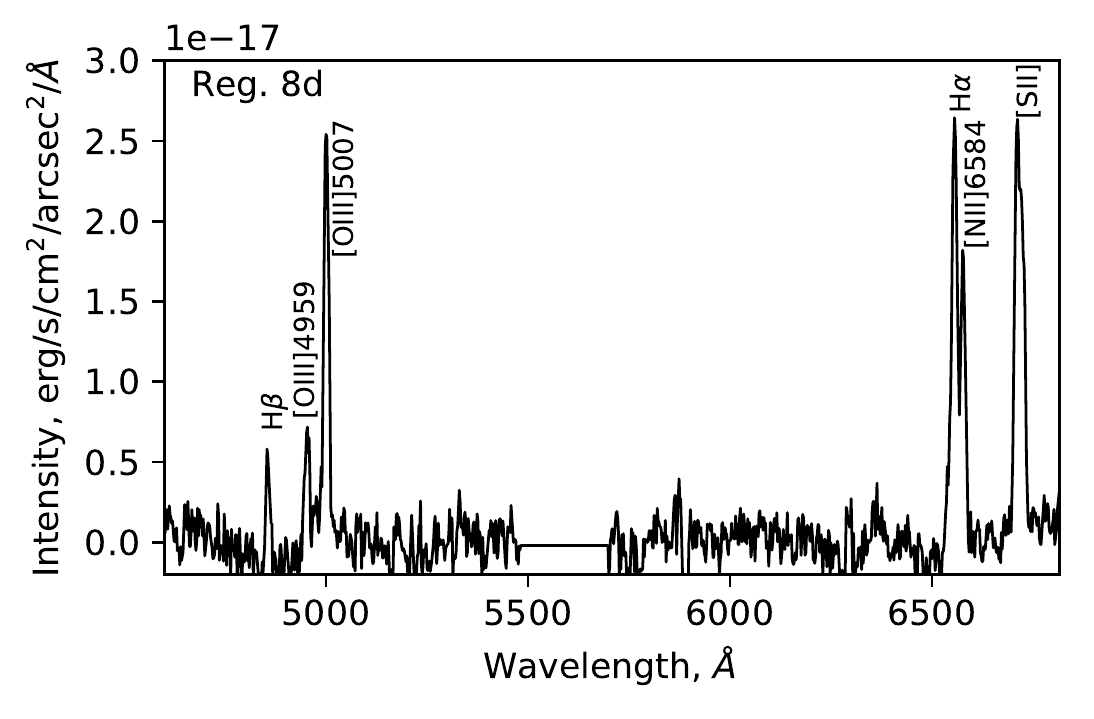}
    \includegraphics[width=0.8\columnwidth]{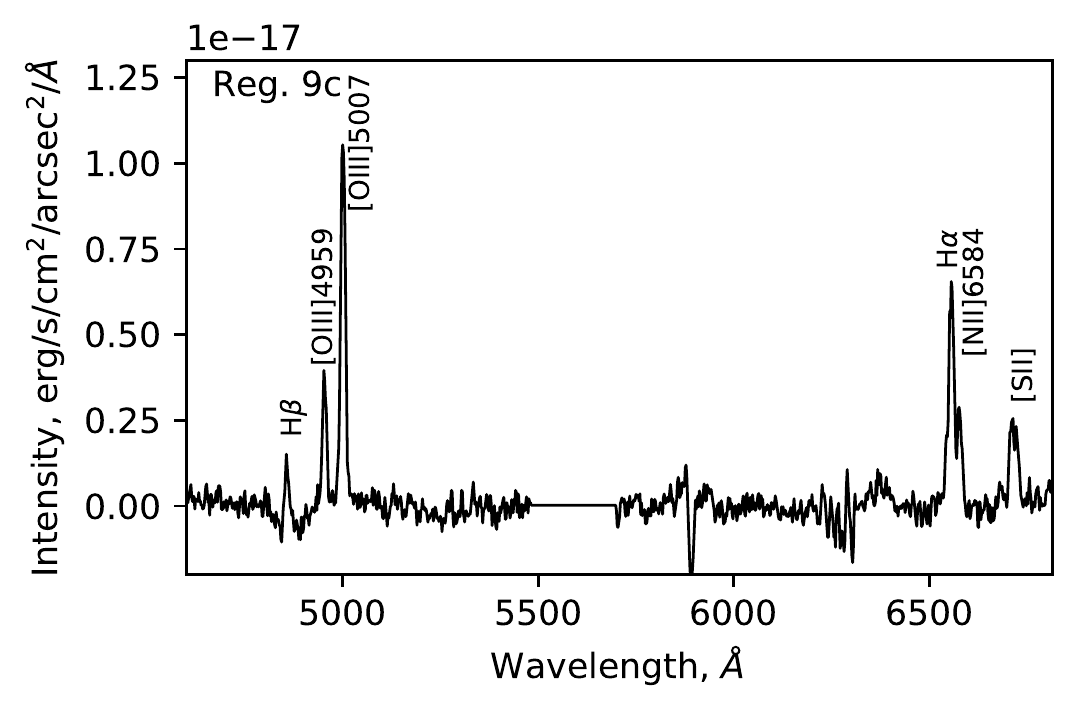}
  \caption{Integrated low-resolution SCORPIO spectra of objects (from top to bottom):  {VIEM} 6, 7, 8d, and 9c.}
              \label{lowresspectra}
\end{figure}

\begin{figure}
 \centering
 \includegraphics[width=0.8\columnwidth]{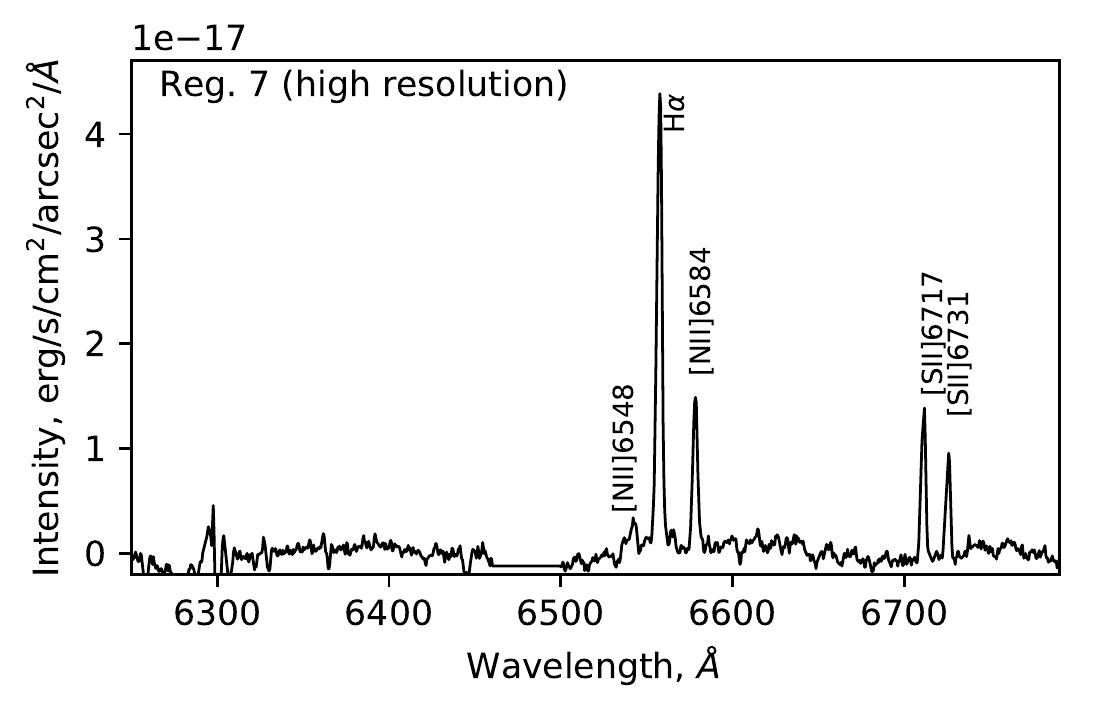} 
 \includegraphics[width=0.8\columnwidth]{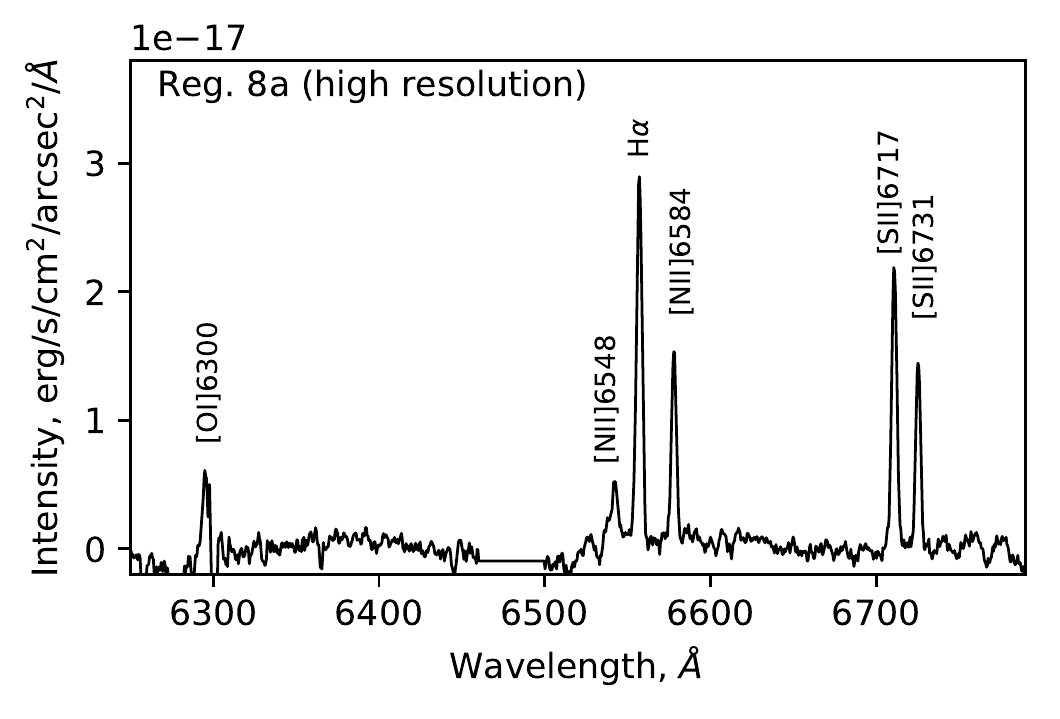}
\caption{Integrated high-resolution SCORPIO spectra of objects  {VIEM} 7 (top) and  {VIEM} 8a (bottom).}
   \label{reg7hrspectra}
\end{figure}

The primary goal of our spectroscopic observation was to obtain good-quality spectra of one previously known shell-like SNR candidate, object  {VIEM} 8, in order to reveal its  physical characteristics, kinematics, and eventually its origin. That is why we used multiple slit positions, both in low- and high-spectral-resolution mode, over the central part of NGC 185. 
Observed emission line fluxes obtained from the integrated spectra for the particularly interesting objects and filaments, namely objects  {VIEM} 6, 7, 8a -- 8d, 9a -- 9c, and 10 are listed in Table \ref{table3}. 
The low-resolution spectra allowed us to estimate the reddening $E(B-V)$ from the Balmer decrement, and from the high-resolution spectra we derived the electron density $n_{\rm e}$ from the \hbox{[S\,{\sc ii}]}6717/6731 line ratio  {\citep{Osterbrock}}. Our plots based on low-resolution fluxes were corrected for extinction, whereas for the high-resolution data we used the observed fluxes. The flux ratios \hbox{[O\,{\sc iii}]}/H$\beta$, \hbox{[S\,{\sc ii}]}/H$\alpha$, and \hbox{[N\,{\sc ii}]}/H$\alpha$ almost do not suffer from the extinction  {due to their closeness in wavelength}, and the correction is less than the estimated observational
uncertainties. Examples of integrated spectra of interesting objects and filaments, namely  {VIEM} 6, 7, 8a, 8d, and 9c, obtained in either low- or high-resolution mode are given in Figs. \ref{lowresspectra} and \ref{reg7hrspectra}. Enhanced \hbox{[S\,{\sc ii}]} emission is clearly seen in all spectra. 

  \begin{figure}
   \centering
   \includegraphics[width=9cm]{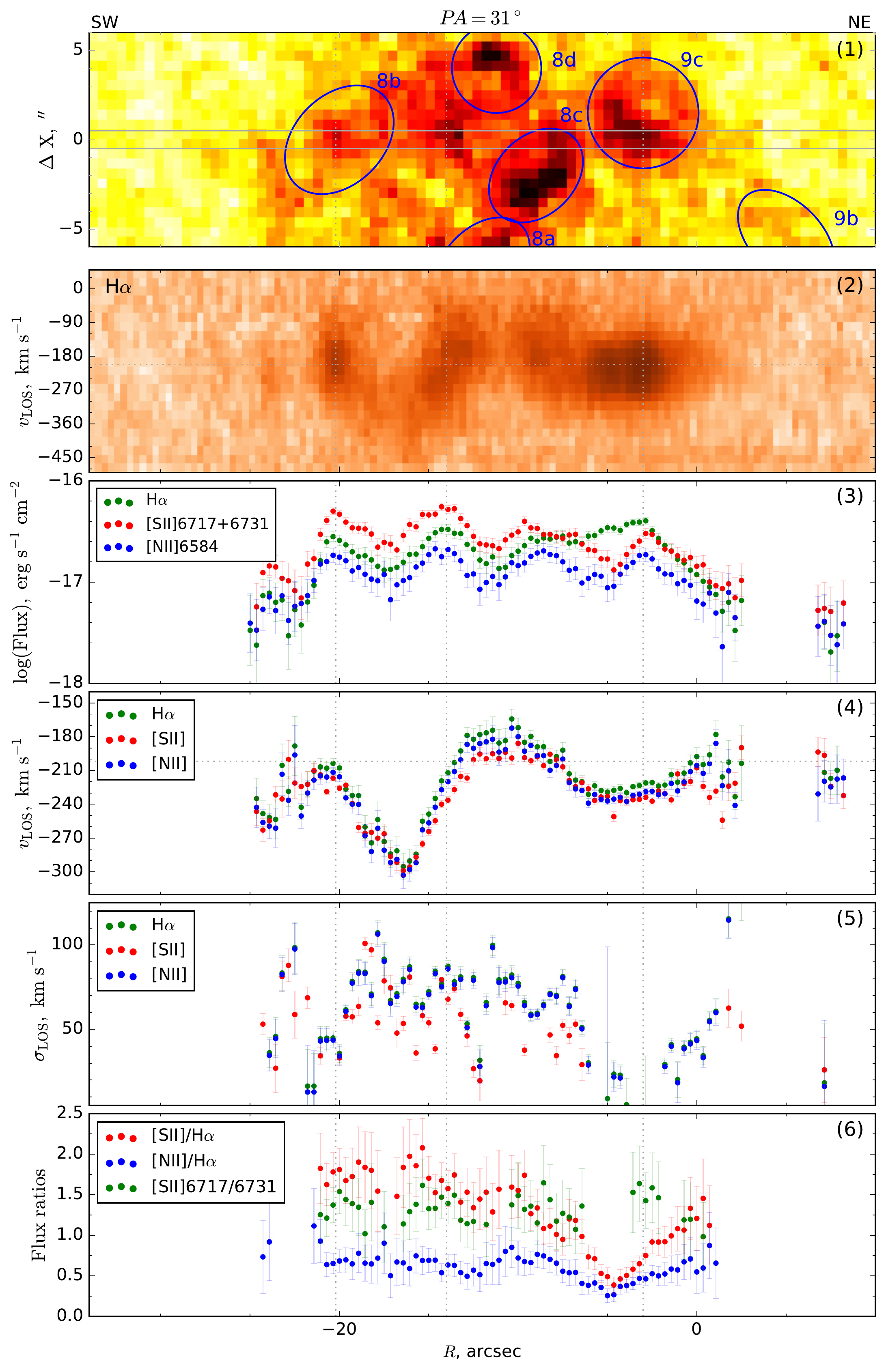}
      \caption{Plots for high-resolution spectra of PA31. The slit is pointed on objects  {VIEM} 8b, 8c and 9c, from the left side. Panels from top: 1) slit position on H$\alpha$ image; 2) PV diagram; 3) flux logarithm in observed emission lines; 4) heliocentric velocity; 5) velocity dispersion (corrected for instrumental profile); 6) flux ratios - \hbox{[S\,{\sc ii}]}6717+6731/H$\alpha$, \hbox{[N\,{\sc ii}]}6584/H$\alpha$ and \hbox{[S\,{\sc ii}]}6717/6731. 
              }
         \label{pa31}
   \end{figure}

\begin{figure}
 \centering
 \includegraphics[width=9cm]{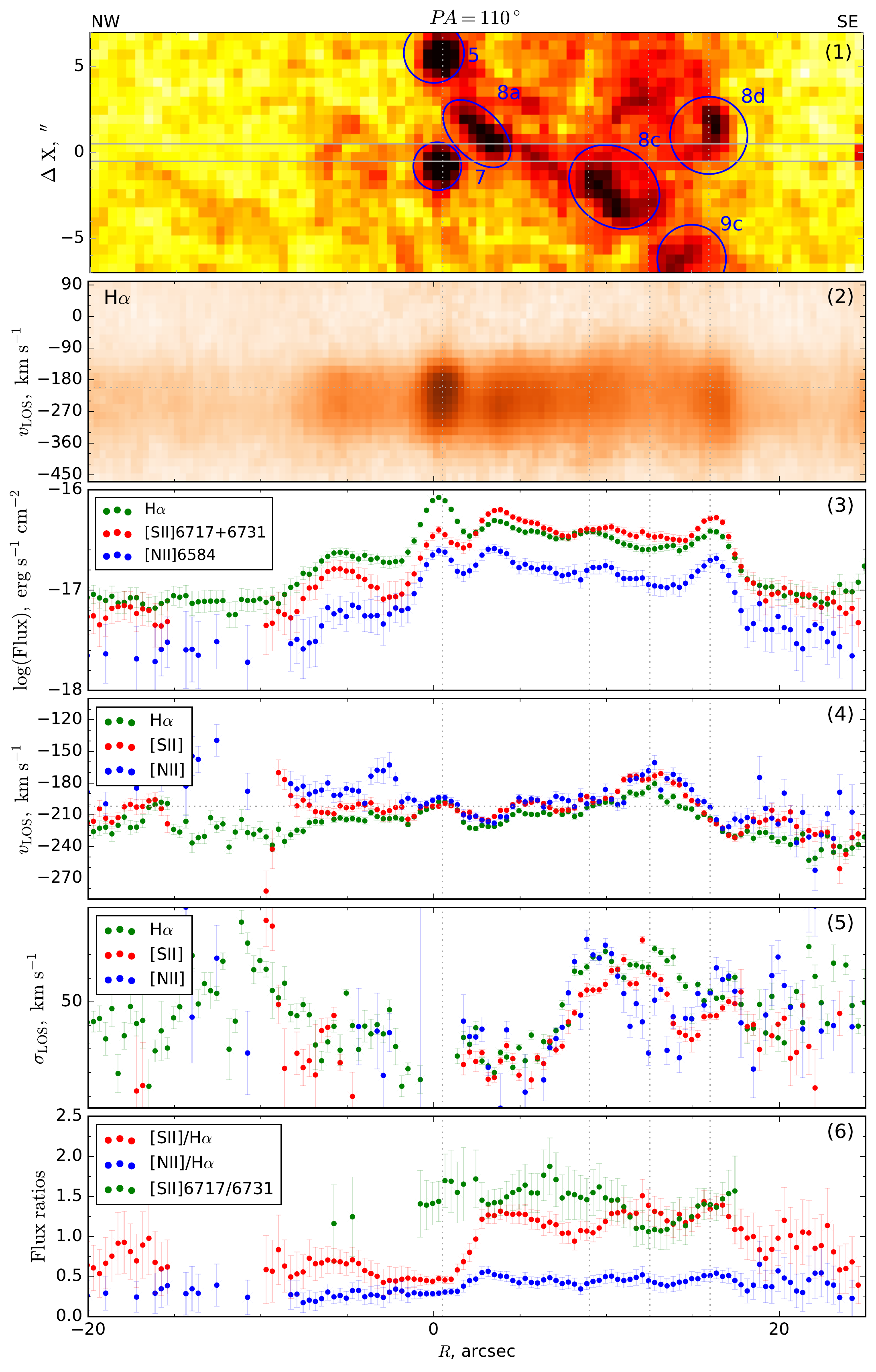}
 \caption{Plots for high-resolution spectra of PA110. The slit is pointed on objects  {VIEM} 7, 8a, 8c, and 8d, from the left side. Panel designations are as in Fig. \ref{pa31}. 
           }
      \label{pa110}
\end{figure}

\begin{figure}
   \centering
   \includegraphics[width=9cm]{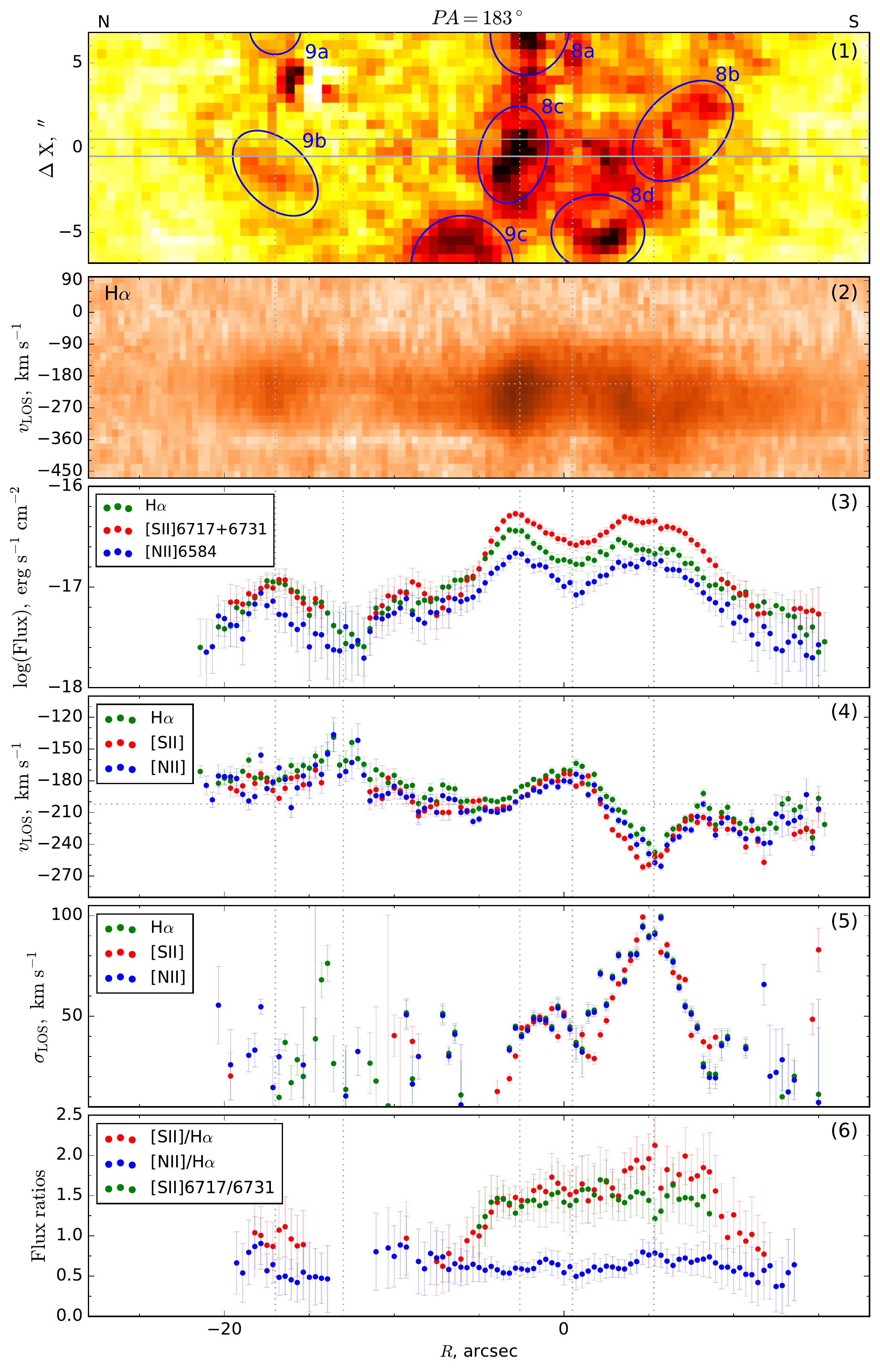}
      \caption{Plots for high-resolution spectra of PA183. The slit is pointed on objects  {VIEM} 9b, 8c, and 8b, from the left side. Panel designations are as in Fig. \ref{pa31}. 
              }
         \label{pa183}
   \end{figure}

Three high-resolution spectra (PA31, PA110 and PA 183) allowed us to analyze line-of-sight (LOS) velocity, velocity dispersion, and flux ratios  {of detected lines} along the slits. Velocities and dispersion {  were measured by fitting a single Gaussian to the observed emission line profiles before} correcting for variations of line-spread function (LSF). Heliocentric correction was also applied.
For these spectra the corresponding Figs. \ref{pa31}, \ref{pa110}, and \ref{pa183} contain six panels:
1) slit position on H$\alpha$ image; 2) position--velocity (PV) diagram; 
3) flux logarithm in observed emission lines;
4) heliocentric line-of-sight velocity;
5) velocity dispersion (corrected for instrumental broadening), and
6) flux ratios - \hbox{[S\,{\sc ii}]}6717+6731/H$\alpha$, \hbox{[N\,{\sc ii}]}6584/H$\alpha,$ and \hbox{[S\,{\sc ii}]}6717/6731.
 {The above panels were selected to support our investigation of the origin (panels 1, 3, 6) and kinematics (panels 2, 4, 5) of the detected emission regions along the slit position. The slit PA31 is crossing objects  {VIEM} 8b, 8c, and 9c (Fig. \ref{pa31}, panel 1), which exhibit very enhanced \hbox{[S\,{\sc ii}]} emission compared to H$\alpha$ (Fig. \ref{pa31}, panels 3, 6). This is confirmed with the other two slit positions, that is, for  objects  {VIEM} 8a, 8c, 8d (Fig. \ref{pa110}, panels 1, 3, 6), and 9b, 8c, 8b (Fig. \ref{pa183}, panels 1, 3, 6). On the other hand, object  {VIEM} 7 has higher brightness (Fig. \ref{pa110}, panel 1), but does not follow the same trend in \hbox{[S\,{\sc ii}]}, which is supported with the observed lower \hbox{[S\,{\sc ii}]}/H$\alpha$ line ratio (Fig. \ref{pa110}, panel 6). We note that the enhanced line emission is seen all across the extended objects  {VIEM} 8 and 9, and not only in the inferred clumps (8a -- 8d, 9a -- 9c), leading us to argue that they are most likely part of the SNR shells.} 

{  Panel 2 in each of the Figs. \ref{pa31}, \ref{pa110}, and \ref{pa183} shows the observed PV diagrams in the H$\alpha$ line, while panel 4 gives the measured LOS velocity in H$\alpha$, \hbox{[N\,{\sc ii}],} and \hbox{[S\,{\sc ii}]} lines. These panels support the SNR origin of objects  {VIEM} 8 and 9, demonstrating the presence of velocity half-ellipse (best seen in panel 2 of Figs. \ref{pa31} and \ref{pa183}) corresponding to shells expanding with high velocities (up to 90 km s$^{-1}$) in an inhomogeneous medium.  In contrast, object  {VIEM} 7 does not show any expansion (Fig. \ref{pa110}, panels 2, 4). The distribution of velocity dispersion (panel 5 in Figs. \ref{pa31}--\ref{pa183}) reveals peaks toward the maximum values of LOS velocity deviation from regular distribution, which is also typical for expanding superbubbles with poorly resolved approaching and receding sides. The measured high-velocity dispersion of $\sim$100 km $^{-1}$ proves the shock-expansion in  {VIEM} 8 and 9, whereas in  {VIEM} 7 the velocity dispersion could not be measured since the line width was lower than the instrumental one, indicating much lower velocity dispersion.}

\begin{figure}
   \centering
   \includegraphics[width=9cm]{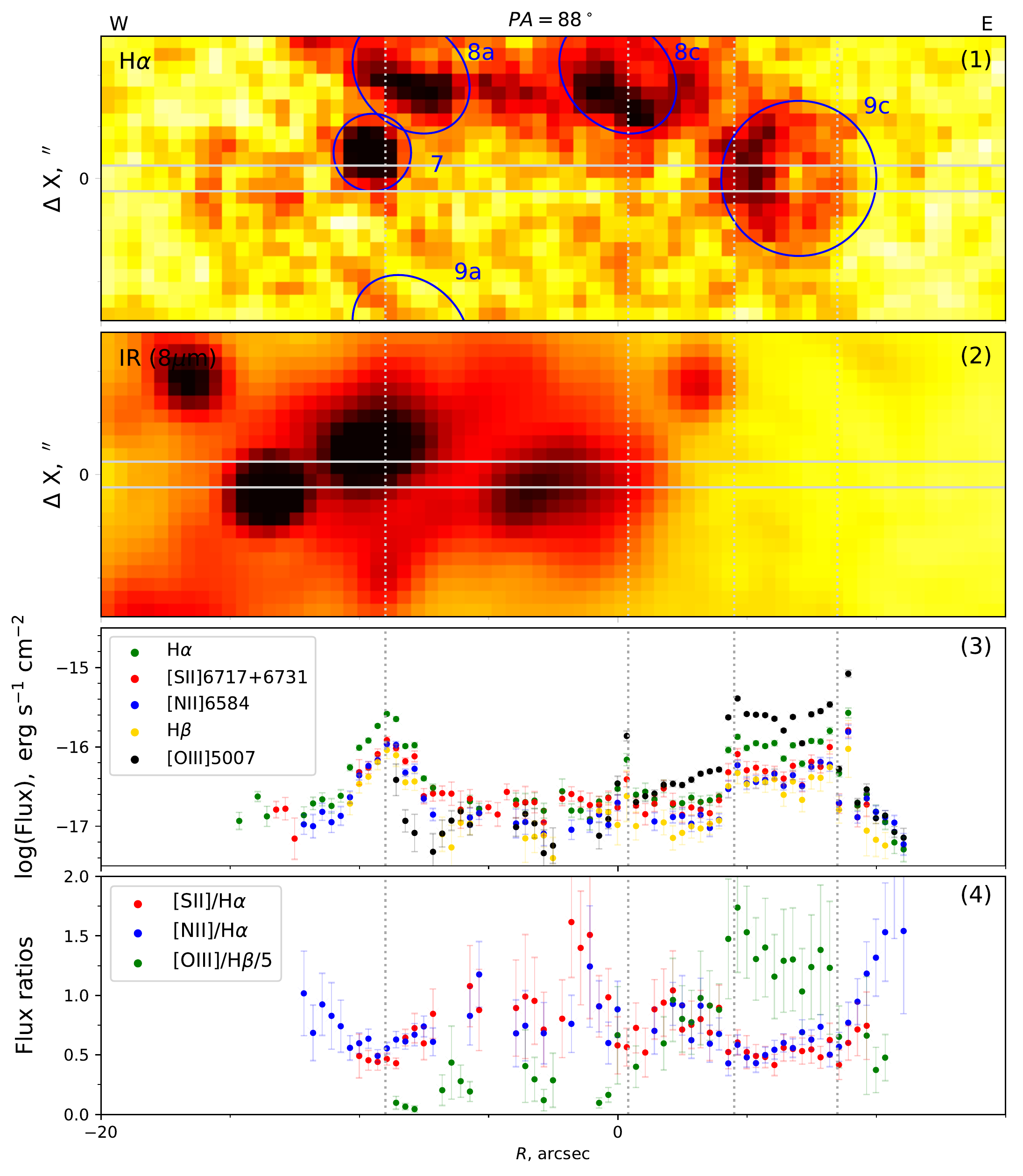}
      \caption{Plots for low resolution spectra of PA88. The slit is pointed on objects  {VIEM} 7 and 9c, from the left side. Panels from top: 1) slit position on H$\alpha$ image;
2) \textit{Spitczer} IRAC 8 $\mu$m map;
3) Flux logarithm in observed emission lines;
4) Flux ratios - \hbox{[S\,{\sc ii}]}6717+6731/H$\alpha$, \hbox{[N\,{\sc ii}]}6584/H$\alpha$ and \hbox{[O\,{\sc iii}]}/H$\beta$ {  (divided by 5 in order to fit the same range as for other flux ratios)}.  
              }
         \label{pa88}
   \end{figure}

\begin{figure}
  \centering
  \includegraphics[width=9cm]{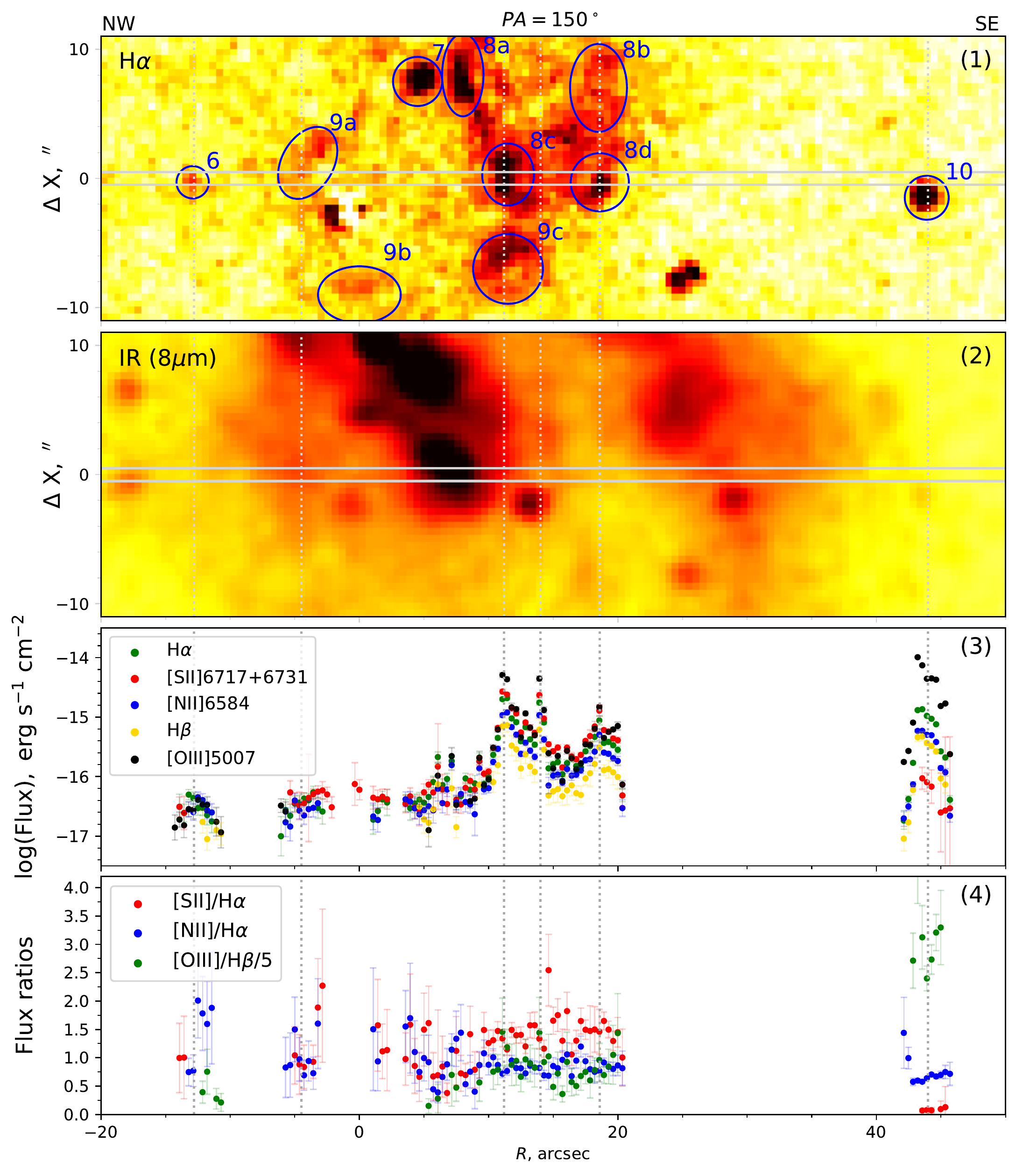}
  \caption{Plots for low-resolution spectra of PA150. The slit is pointed on objects  {VIEM} 6, 9a, 8c, 8d and 10, from the left side. Panel designations are as in Fig. \ref{pa88}. 
            }
       \label{pa150}
   \end{figure}

Low-resolution spectra (PA88 and PA150) did not have enough resolution to analyze gas kinematics or \hbox{[S\,{\sc ii}]}6717/6731 lines ratio, and therefore we did not include the corresponding panels  {(2, 4, 5)}
in  Figs. \ref{pa88} and \ref{pa150}.  {However, these panels are important since they cover the blue part of the spectrum and the slit crosses objects  {VIEM} 6, 7, 8c, 8d, 9a, 9c, and 10 (Figs. \ref{pa88} and \ref{pa150}, panel 1). In addition to the enhanced \hbox{[S\,{\sc ii}]} emission in the spectra of these objects (Fig. \ref{pa88}, panel 4), there is also enhanced \hbox{[O\,{\sc iii}]} emission, which is absent in  {VIEM} 7 (Fig. \ref{pa88}, panels 3, 4), and present but much lower than \hbox{[S\,{\sc ii}]} and H$\alpha$ emission in  {VIEM} 8 and 9 (Figs. \ref{pa88} and \ref{pa150}, panels 3, 4).} 
We added a panel with \textit{Spitzer} IRAC 8 $\mu$m maps to  {trace} warm dust emission in the central part of NGC 185 (Figs. \ref{pa88} and \ref{pa150}, panels 2) . Some dust condensations are coincident with the positions of  {VIEM} 7 and the central part of  {VIEM} 9. 

\begin{figure}
   \centering
   \includegraphics[width=9cm]{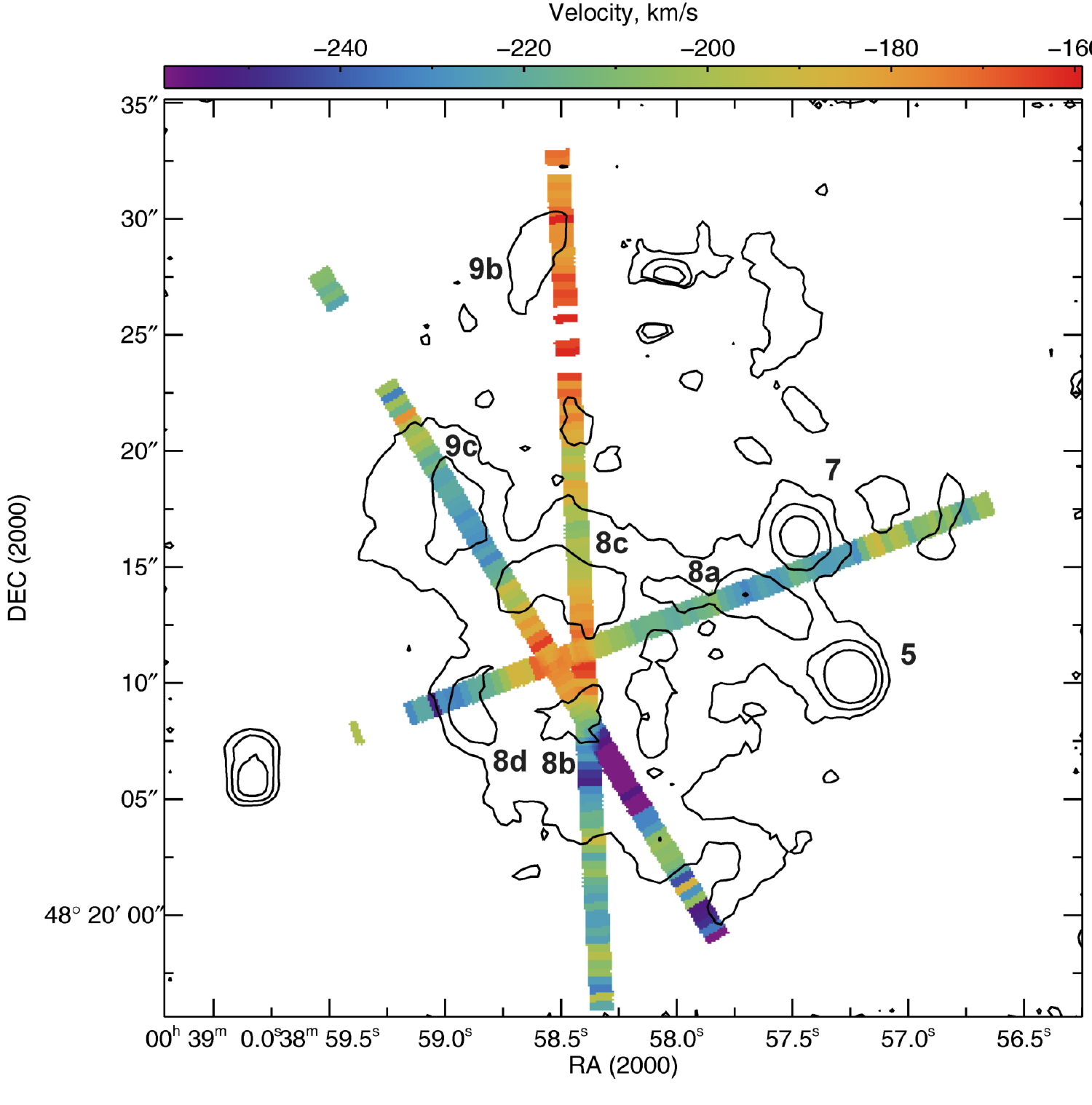}
      \caption{{  Two-dimensional} velocity map {  in the H$\alpha$ line} of the central $40\arcsec \times 40\arcsec$ of NGC 185, constructed from  {three slit positions for} high-resolution spectra, {  with objects denoted as in Fig. \ref{n185-CII}}. Contours  denote H$\alpha$ intensity  {from narrowband imaging. Assuming that heliocentric velocity of the galaxy is $-203.8$~km~s$^{-1}$ \citep{McConnachie2012}, we observe the receding side of shells  {VIEM} 8 and  {VIEM} 9.}
      }
         \label{velmap}
   \end{figure}

 {Finally,} from three slit positions in high-resolution spectroscopic mode we were able to construct a  {partial} line-of-sight velocity map of the central region of the galaxy (see Fig. \ref{velmap}). {  This map was constructed as follows. For each pixel crossed by the slit, the velocity {was derived from}  H$\alpha$ line, as given in panel 4 of Figs.~\ref{pa31}--\ref{pa183}. The width of the strips in Fig. \ref{velmap} is equal to the width of the slits (1 arcsec). In the regions where several slits overlap, we use the mean value of LOS velocity. We note that the measured velocities at the same positions in different slits are in agreement within the uncertainties. Contours in Fig. \ref{velmap} trace H$\alpha$ photometric intensities.}

As follows from Fig.~\ref{velmap}, the central part of the shells of objects    {VIEM} 8 and  {VIEM} 9 are redshifted relative to the heliocentric velocity of NGC~185  ($-203.8$~km~s$^{-1}$ according to \citealt{McConnachie2012}). At the same time, the southern part of  {VIEM} 8 demonstrates a blueshifted velocity (region 8b).  Both objects  {VIEM} 8 and  {VIEM} 9 are very clumpy, that is, they contain substructures (see Fig. \ref{n185-CII}) which are expanding in highly inhomogeneous ISM. That is probably why we do not  clearly observe both approaching and receding sides of the expanding shells. Observing only the receding side of an expanding shell could be explained if higher-density gas condensations, which are observed in the central part of this galaxy (see  \hbox{[C\,{\sc ii}]} contours in Fig. \ref{n185-CII}), are located on the far side of these objects, enhancing the emission. 

\subsection{Diagnostic diagrams}
\begin{figure*}
   \centering
   \includegraphics[width=17.4cm]{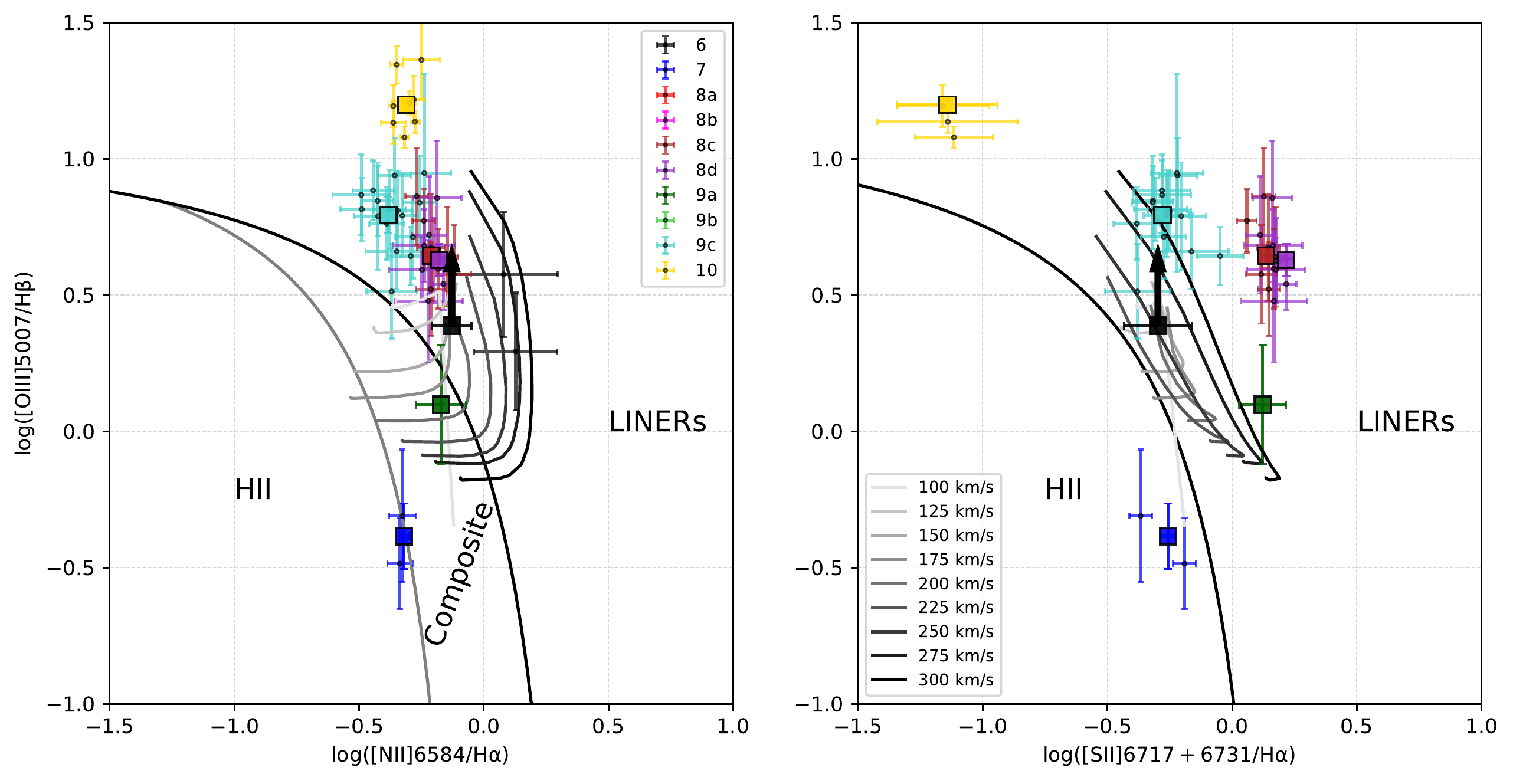}
   
   \caption{Diagram showing BPT plots \hbox{[O\,{\sc iii}]}/H$\beta$ versus \hbox{[N\,{\sc ii}]}/H$\alpha$ (left) and \hbox{[O\,{\sc iii}]}/H$\beta$ versus \hbox{[S\,{\sc ii}]}/H$\alpha$ (right) with overlaid
\citet{Kewley2001} and \citet{Kauffmann2003} separation lines (black and gray, respectively) and MAPPINGS III \citep{Allen2008} radiative shock models for solar abundances and preshock electron density $n_{\rm e}=10$ cm$^{-3}$. Each pixel along the slit is plotted with a filled circle with error bars. Large squares show results from spectra integrated over the larger area (usually the whole part of the region crossed by the slit). Colors correspond to those pixels lying within the borders of  objects. For consistency with colors used in Fig. \ref{diag}, all objects are listed, whereas BPT plots are constructed from
only low-resolution spectra which cover the large spectral range -- objects  {VIEM} 6, 7, 8c, 8d, 9a, 9c, and 10. }
              \label{bpt}
    \end{figure*}

To identify the source of gas ionization in each object, in Fig. \ref{bpt} we plotted BPT diagnostic diagrams obtained from low-resolution spectra only, since they cover  {both the blue and red part of the optical spectrum}.  The BPT plots of \hbox{[O\,{\sc iii}]}/H$\beta$ versus \hbox{[N\,{\sc ii}]}/H$\alpha$ (Fig. \ref{bpt}, left) and \hbox{[O\,{\sc iii}]}/H$\beta$ versus \hbox{[S\,{\sc ii}]}/H$\alpha$ (Fig. \ref{bpt}, right) are overlaid with \citet{Kewley2001} and \citet{Kauffmann2003} separation lines to distinguish regions with different excitation mechanisms. 

\begin{figure}
   \centering
   \includegraphics[width=9cm]{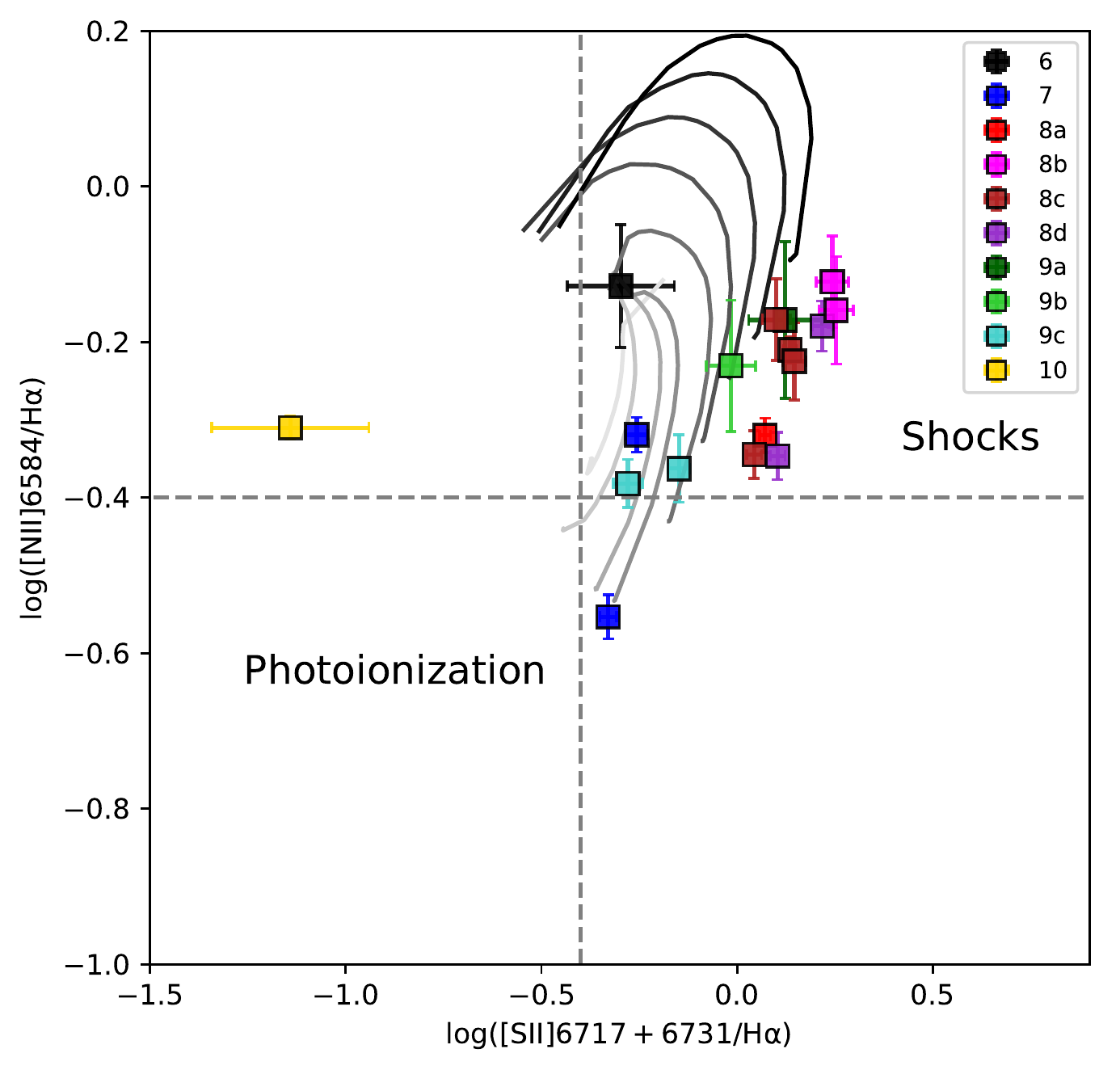}
   
   \caption{H$\alpha$/\hbox{[N\,{\sc ii}]} versus H$\alpha$/\hbox{[S\,{\sc ii}]} diagnostic diagram. Based on the discussion in \citet{Stasinska2006}, simple separation lines are plotted to indicate regions with different ionization mechanisms: the bottom left quadrant is occupied by photoinized objects, while the upper right quadrant is occupied by shock-heated objects. MAPPINGS III \citep{Allen2008} radiative shock models for solar abundances and preshock electron density $n_{\rm e}=10$ cm$^{-3}$ are overlaid. }

              \label{diag}
    \end{figure}

Also, we constructed a  \hbox{[N\,{\sc ii}]}/H$\alpha$ versus \hbox{[S\,{\sc ii}]}/H$\alpha$ diagnostic diagram (Fig. \ref{diag}), showing the flux ratios of lines with different excitation mechanisms and similar wavelengths, to avoid their dependence on dust extinction. Figure \ref{diag} shows data for the same objects observed in both low- and high-resolution mode.  {We note that since} slit positions are crossing different parts of extended regions, different line ratios  {could be measured} for the same objects visible in Fig. \ref{diag}. Here the dashed lines are in accordance with the discussion presented in \citet{Stasinska2006}, which showed that in the case of ionization by young stars,  line ratios follow the relations log\hbox{[S\,{\sc ii}]}/H$\alpha$  $<-0.4$ and log\hbox{[N\,{\sc ii}]}/H$\alpha$ $<-0.4$. A higher relative intensity of forbidden lines means that shocks, or even a nonthermal source (AGN), contribute appreciably to the ionization. 

On all diagnostic diagrams we plot MAPPINGS III \citep{Allen2008} radiative shock models for solar abundances and pre-shock electron density $n_{\rm e}=10$ cm$^{-3}$  {to indicate the source of ionization}. Our density measurements suggest values of  $n_{\rm e}\sim100$ cm$^{-3}$, but those are values in the shell, where outer gas has previously been compressed by the shock. Although NGC~185 has lower metallicity, only models for solar abundances are available for different density assumptions, and  {that is why we used higher metallicity than suspected for this galaxy.}

Before interpreting the positions of data points on the diagram, we reiterate the following: (i) the boundaries of the photoionization domain are controversial (i.e., the boundaries are determined not only by particular models, but also by the statistics of observations of different galaxy samples); (ii) in real objects, the combined effect of several ionization sources is present. Thus, according to \citet{Stasinska2006}, the $-0.2 <$ log\hbox{[N\,{\sc ii}]}/H$\alpha$ $< -0.4$ region corresponds to a composite source of ionization (photoionization by stars plus the effect of some shocks), and shocks dominate only for log\hbox{[N\,{\sc ii}]}/H$\alpha$ $>-0.4$.

\subsection{Archival  {X-ray and radio} data results}

\begin{figure}
\centering
\includegraphics[width=9cm]{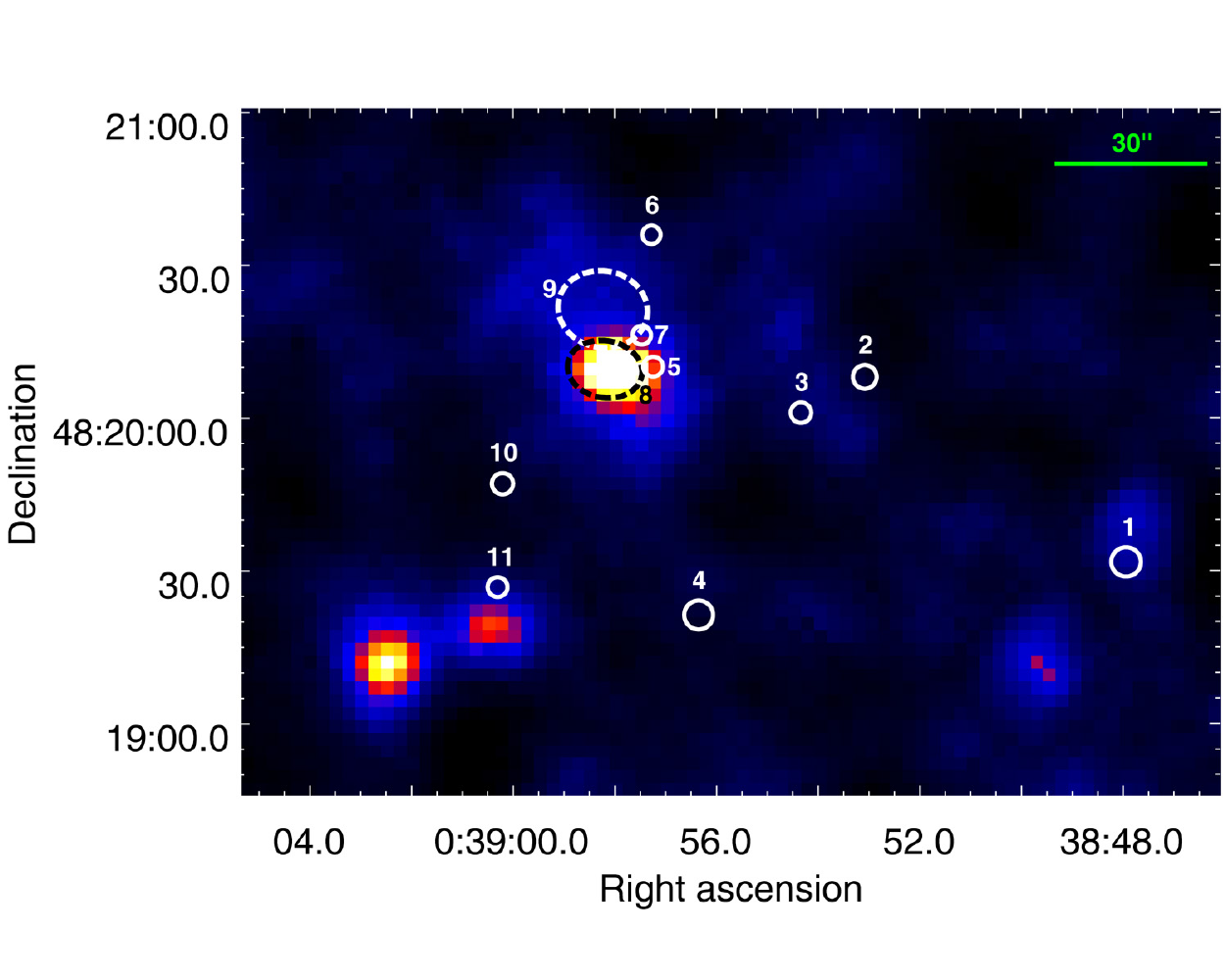}
\caption{\footnotesize{Adaptively smoothed, exposure-corrected combined EPIC image of the central parts of NGC 185. The image depicts emission detected over the 0.4 keV to 7.0 keV energy range (scaled with zscale/sinh and 100 smoothing counts). The optically detected sources are represented with filled circles, except objects  {VIEM} 8 and 9 which are indicated with dashed circles.}}
\label{xray1}
\end{figure}

In Fig. \ref{xray1} we present an adaptively smoothed, exposure-corrected combined \textit{XMM-Newton} EPIC image of the central part of NGC 185. 
The optically detected sources are represented by filled circles, except objects  {VIEM} 8 and 9 which are shown with dashed circles. It is clearly visible that there is an X-ray source that is mainly coincident with the optical position of object  {VIEM} 8. We have estimated the hardness ratios for the source region, as described in \citep{SHP13}. 
Based on the images in different energy bands and {according to its hardness hardness ratio}, we conclude that VIEM 8 is a soft X-ray source. The obtained results, $\mathrm{HR_{1}}>0$ and $\mathrm{HR_{2}}<0$, are in accordance with the conclusions stated in \citet{SHH18} for SNRs in the northern disk of the galaxy M31. Soft X-rays indicate a thermal origin, as expected for an SNR, and enhance the likelihood that are detecting emission from a remnant. We would also like to stress that our results agree very well with the \textit{XMM-Newton} Serendipitous Source Catalog (3XMM DR8 Version; \citealt{3XMM}), that reported an extended X-ray source 3XMM J003858.1+482010; ($\mathrm{HR_{1}}=0.48$, $\mathrm{HR_{2}}=-0.47$)  with a  radius of $7\arcsec$.
The EPIC total band flux is of 5.33$\times$10$^{-15}$ erg cm$^{-2}$ s$^{-1}$ in the 0.2-12.0 keV energy range. 
Finally, we would like to stress the fact that the center of the NGC 185 galaxy itself is projected to the same part of the sky as the optical object  {VIEM} 8 and the identified X-ray source. In that sense, any association between the soft X-ray source detected by {\it XMM-Newton} and an SNR identified at optical wavelengths must be viewed only as indicative. In the future, new deeper observations, for example with the {\it Chandra} X-ray Observatory, which has the necessary angular resolution, could help in resolving the diffuse and discrete X-ray source population (including X-ray sources associated with SNRs) seen toward the nucleus of NGC 185.

Finally, a simple analysis of the X-ray spectrum of the source 3XMM J003858.1+482010 
gives $kT=0.18^{\ \ 0.07}_{-0.04}\ \mathrm{keV}$ and $N_{\rm H, int}=0.38^{\ \ 0.17}_{-0.32}\times10^{21}\ \mathrm{cm^{-2}}$ ($1\sigma$ range estimate for all parameters). We also tried to fit the spectrum with single-temperature and single-ionization-age \verb"NEI" plasma, which in this case requires an ionization timescale higher than $10^{12}\ \mathrm{s\ cm^{-3}}$, indicating CIE plasma. Both models require rather high intrinsic absorption in NGC 185. On the other hand, the model that incorporates a simple sum of the \verb"APEC" model and a power-law spectrum, gives $kT=0.27^{\ \ 0.05}_{-0.06}\ \mathrm{keV}$ and a power-law photon index of $\Gamma=2.94^{\ \ \!1.00}_{-1.01}$, with negligible intrinsic absorption. This may imply the presence of an unseen compact object. {The model most closely resembles reality when a CIE plasma is incorporated.}  

\begin{figure}
\centering
\includegraphics[width=9cm]{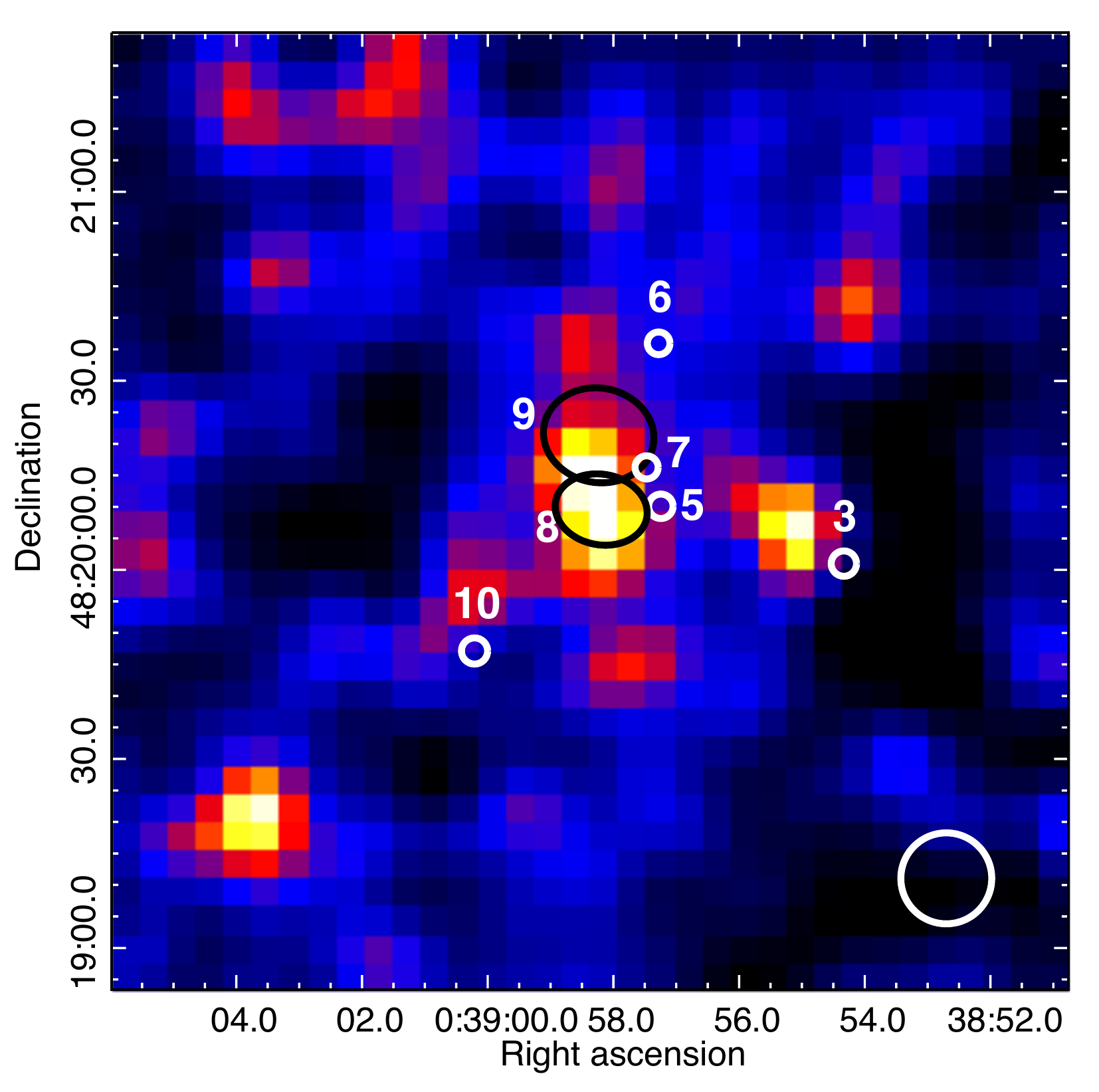}
\caption{\footnotesize{Archival VLA image of NGC 185 central region of the size 2.5\arcmin $\times$ 2.5\arcmin \ (beam size of 14.4\arcsec indicated in the bottom left).  The optically detected objects are represented by filled circles.}}
\label{vla}
\end{figure}

{The central  2.5\arcmin $\times$ 2.5\arcmin of NGC 185 observed with VLA is shown in Fig. \ref{vla}, with the positions of optically detected objects. There is some indication of the diffuse radio continuum emission coincident with objects  {VIEM} 8 and 9, however the resolution is not good enough to resolve the complex  emission in the center. We obtained a robust flux estimate for the aperture size of 15\arcsec \ centered on the position of  {VIEM} 8, and obtained the flux of 1.4 mJy. The corresponding radio luminosity at the distance of NGC 185 is 1.4$\times$10$^{34}$ erg/s, which is comparable to the average radio luminosity of typical Galactic shell-like radio SNRs.}

In addition,  the flux of 1.4 mJy corresponds to a surface brightness of $0.35 \times 10^{-20}$ W m$^{-2}$Hz$^{-1}$sr$^{-1}$.  This is above the surface brightness detection limit for samples of extragalactic SNRs, for example the limit for the Large Magellanic Cloud (LMC) sample is $0.03 \times 10^{-20}$ W m$^{-2}$Hz$^{-1}$sr$^{-1}$, which was detected with a different instrumental setup \citep{Bozzetto2017}.  The recently published deep  Jansky VLA (JVLA) radio survey of M33 \citep{White2019} obtained radio emission for 155 of the 217 optical SNRs in this galaxy, reaching a limiting sensitivity of 20 $\mu$Jy and a resolution of 5.9\arcsec, which corresponds to an even lower surface brightness limit than in the LMC: $0.02 \times 10^{-20}$ W m$^{-2}$Hz$^{-1}$sr$^{-1}$. Therefore, new radio observations with higher spatial resolution and better sensitivity may resolve radio emission corresponding to detected optical SNRs or other emission nebulae in the central part of NGC 185. 

\subsection{Discussion on individual objects}

Here we discuss in detail  detected objects that were classified as optical SNRs, and two more objects that show signs of shock-ionization.

\textit{ \bf{VIEM 8.}} Figure \ref{n185-CII} clearly shows the partial shell of a shock-heated object,  {VIEM} 8, which corresponds to a  SNR previously
detected by \citet{Gallagher1984} and \citet{MartinezDelgado1999}, and has a \hbox{[S\,{\sc ii}]}/H$\alpha$ photometric ratio of 1.24 (as in \citealt{Gallagher1984}) and a diameter of 15\arcsec= 45 pc. For the same distance, \citet{MartinezDelgado1999} estimated a diameter of 80 pc and a flux of $30\pm 3\times 10^{-15}$  erg cm$^{-1}$ s$^{-2}$, which was in agreement with \citet{YoungLo1997}. Our estimated H$\alpha$ flux for this object, $28.7\pm5.0 \times 10^{-15}$ erg cm$^{-1}$ s$^{-2}$, also agrees with previous assessments. With deep enough exposures we obtained good spectra on several positions of the shell-like structure, object  {VIEM} 8 (see Fig. \ref{slits} for slit positions). Spectroscopic \hbox{[S\,{\sc ii}]}/H$\alpha$ ratios for this extended object are in the range of 1.4 -- 2.0 (depending on position) confirming the SNR origin,  which again agrees  with the Gallagher et al. (1984) value of \hbox{[S\,{\sc ii}]}/H$\alpha$ = 1.2 $\pm$ 0.3. \hbox{[N\,{\sc ii}]}/H$\alpha$ line ratios are in the range of 0.6 -- 0.75, typical for radiative SNR spectra.
From the ratio of \hbox{[S\,{\sc ii}]}$\lambda$6717/\hbox{[S\,{\sc ii}]}$\lambda$6731 lines we obtained an electron density of $\sim$ 100 cm$^{-3}$ for  {VIEM} 8 (Table \ref{table3} and Figs. \ref{pa31},\ref{pa110} and \ref{pa183},  {panels 6}). From diameter and density, we can asses the age of the remnant. Ages of SNRs are usually obtained assuming the Sedov-Taylor adiabatic stage of SNR evolution. Furthermore, the radius of the shock front $R_s$ is given by \citep{Cox1972}
\begin{equation} \label{eq1}
R_s=12.9 \, t{_{4}}^{2/5} \,  \left ( {\frac{E_0}{n_{0}}} \right )^{1/5}\rm{pc},
\end{equation}
where the age of the remnant  $t_4$ is in multiples of $10^4$ yr, $E_0$ is the initial energy in multiples of $10^{51}$ erg, and $n_0$ is ambient density in cm$^{-3}$. For a radius of 22.5 pc, $E_0=1$ and  $n_0=10$, and therefore  {VIEM} 8 would be $1.1 \times 10^5$ yr old. This age is only a lower limit. The time of transition to radiative phase $t_{\rm{tran}}$, given by \citet{Bandiera2004} is  
 \begin{equation} \label{eq2}
t_{\rm{tran}}=2.84 \, 10^4 \, E_{0}^{4/17}n_{0}^{-9/17}.
\end{equation}
For the assumed parameters of   {VIEM} 8, $t_{\rm{tran}}=0.8\times 10^4$ yr, indicating that this SNR candidate would be in the late radiative phase.
Object  {VIEM} 8 shows filaments of enhanced emission (8a -- 8d), which are probably the result of an inhomogeneous density  of the surrounding medium, as indicated from the distribution of warm dust  (see Figs. \ref{pa88} and \ref{pa150},  {panels 4}).
 
 {From diagnostic diagrams we see that different filaments of object  {VIEM} 8 (8a -- 8d) occupy the same, shock-ionized regions on both Figs. \ref{bpt} and \ref{diag}, and this is  {an additional} reason for why we propose that they have the same origin, being parts of the radiative SNR. }
 
 {The H$\alpha$ PV diagram (Fig. \ref{pa183}, panel 2)  shows the velocity half-ellipse for the region  {VIEM} 8b, which is expected for an expanding shell}. {If we adopt the highest redshifted velocity observed toward the center of the   {VIEM} 8 shell (see position $\sim -17$~arcsec in Fig.~\ref{pa31},  {panel 4}, and $\sim 11$~arcsec in Fig.~\ref{pa183},  {panel 4}) as the velocity of the expansion observed from the receding side, then we obtain an estimate of the expansion velocity of $V_{\rm exp} \sim 90$~km~s$^{-1}$. As follows from Eq.~\ref{eq1}, $V_{\rm exp}=0.4R_S/t_6$ at the Sedov-Taylor phase of an SNR. This gives us an estimate of the kinematical age of  {VIEM} 8 : $t \sim 1\times10^5$ years, which is in agreement with the previously  obtained value from its radius, assuming the explosion energy $E_0=10^{51}$~erg and density $n=10$ cm$^{-3}$. }

 {In addition, the optical SNR, object  {VIEM} 8, has its unresolved counterparts in the X-ray and radio-emission (see Section 5.4), adding to the evidence of its nature as an SNR. {{According to its large diameter and small expansion velocity, we propose that VIEM 8 is an evolved radiative SNR. Evolved SNRs are expected to have low X-ray and radio emission, which is observed for VIEM 8.}}  

\textit{ {  \bf{VIEM} 9.}} In addition to one possible optical SNR (VIEM 8), our photometric and spectroscopic observations reveal for the first time an extremely faint object,  {VIEM} 9, in our target list, potentially an additional old radiative SNR. Its filaments, denoted  {VIEM} 9a, 9b, and 9c, have also been included within spectroscopic slits, in order to recover as much information as possible about its nature. VIEM 9 exhibits \hbox{[S\,{\sc ii}]}/H$\alpha$ ratios in the range $0.7 - 1.2$, and \hbox{[N\,{\sc ii}]}/H$\alpha$ in the range $0.5 - 1$  {(see Figs. \ref{pa31}, \ref{pa183}, panel 6)}. Region  {VIEM} 9c shows the strongest \hbox{[O\,{\sc iii}]} line, with \hbox{[O\,{\sc iii}]}/H$\beta$=7, while  {VIEM} 9a has fainter \hbox{[O\,{\sc iii}]} lines  {(see Figs. \ref{pa88}, \ref{pa150}, panels 3 and 4)}. According to its diameter of 50 pc, this SNR candidate should be in the radiative phase, aged $\sim6 \times10^5$ years  {(Eq. \ref{eq1})}. {If we assess the expansion velocity of   {VIEM} 9 from Figs. \ref{pa31}  {or \ref{pa183} (panel 4)} in the same way as for  {VIEM} 8,  we obtain $V_{\rm exp}\sim 30$~km~s$^{-1}$. This leads to a kinematical age of $t\sim 3.5\times10^{5}$~years, twice lower than the estimate obtained in from its diameter. This disagreement in age of an SNR may indicate that the real energy of explosion is lower than the adopted one, $10^{51}$ erg, or that the expansion velocity of the shell  of  {VIEM} 9 is lower. The latter is supported by the observed underlying velocity gradient from  {VIEM} 8c to 9b in Fig.~\ref{pa183}  {(panel 4).}}

Some parts of  {VIEM} 9 might be projected on the position of  {VIEM} 7, and also  its southern part interacts with or is projected onto the northern part of  {VIEM} 8 (8a, 8c). {On BPT diagrams (Figs. \ref{bpt} and \ref{diag}), we see some displacement of region 9c toward a photoionization region. It could be that there is an overlap with some background photoinized gas, which is also suggested by our photometric observations. }

\textit{ \bf{VIEM 7.}} This object seems to coincide with object 9-SNR1 from \citet{Goncalves2012}. The observations of these latter authors suggest that   {VIEM} 7 has a diameter of 2 pc, and they claimed that it is an SNR detected by \citet{Gallagher1984}. For this object they measured \hbox{[S\,{\sc ii}]}/H$\alpha$ = 0.53, and our spectroscopic values are in the range  of $0.5-0.6$, depending on the slit position  {(Fig. \ref{pa110}, panel 6)}. From the \hbox{[S\,{\sc ii}]}$\lambda$6717/\hbox{[S\,{\sc ii}]}$\lambda$6731 line ratio, we obtain an electron density of $<200\pm 100$cm$^{-3}$, which is in agreement with the value of  \citet{Goncalves2012}  of $\sim$300 cm$^{-3}$. As already mentioned, \citet{Goncalves2012} did not detect any \hbox{[O\,{\sc iii}]} emission in the spectrum of  {VIEM} 7, and so they concluded that the shock velocity of the SNR is less than 85 km s$^{-1}$, {which is expected for old, evolved SNRs}. These latter authors further suggested that the bright emission-line object (VIEM 7) might be a central part of an old SNR, since it is located approximately in the center of a faint arc-like structure. We question this scenario, since in our image (Fig. \ref{n185-CII}), the arc-like structure (VIEM 8) forms almost a full shell, typical for an evolved SNR, while  {VIEM} 7 is located on the outer part of this shell.  \citet{Vucetic2016}, who performed narrowband imaging, speculated that  {VIEM} 7 could be an additional, young SNR in view of its compact size and different \hbox{[S\,{\sc ii}]}/H$\alpha$ ratio from the rest of the shell-like SNRs. If we assume that  {VIEM} 7 is an SNR in adiabatic phase, with a diameter of 6 pc (upper limit), we obtain an age of $\sim$700 years, assuming that it is evolving in the surrounding density of 10 cm$^{-3}$. {VIEM 7 does not display any significant deviations of the LOS velocity from the heliocentric velocity  {(see Fig. \ref{pa110}, panel 4)}, and therefore we cannot examine its expansion. Also, the velocity dispersion there is lower than the instrumental broadening, and therefore we cannot measure the expansion  {(see Fig. \ref{pa110}, panel 5)}. If we assume that  {VIEM} 7 is an SNR in Sedov-Taylor phase and that the previously published estimates of its age are correct, then it should have an expansion velocity of $V_{\rm exp} \sim 1700$~km~s$^{-1}$ which should definitely  be observed with our spectral data. These facts  {strongly} contradict the possibility that VIEM 7 is a young SNR.}

Therefore, we suggest  {two} possible scenarios:  {VIEM} 7 could be  (i) a compact \hbox{H\,{\sc ii}} region with signatures of overlapping shock-ionized gas from objects  {VIEM} 8 or  {VIEM} 9 (or both), or (ii) part of the old evolved SNR, object  {VIEM} 8 or  {VIEM} 9, which has encountered an ISM condensation of higher concentration at that position,  { slowing down shock propagation, and therefore lowering the \hbox{[S\,{\sc ii}]}/H$\alpha$ ratio}. 
The first scenario is strongly supported by the measured line ratios, which are very different from those of  objects  {VIEM} 8 and  {VIEM} 9, and by the fact that the position of  {VIEM} 7 clearly stands out on the BPT diagrams  {(Fig. \ref{bpt})}.  {The position of  {VIEM} 7 is moved toward the intermediate or photoionized
region  on all three diagnostic diagrams, far from the area occupied by PNe, indicating that it could be an \hbox{H\,{\sc ii}} region.} 
Further, this scenario is strengthened by the suggested presence of a young stellar population and molecular gas by  \citet{MartinezDelgado1999} and \citet{Marleau2010}. {The enhanced \hbox{[S\,{\sc ii}]} emission could be the signature of the overlapping shock-ionized gas.}
The second scenario is supported by the fact that the peak of \hbox{H\,{\sc i}}, CO, and \hbox{[C\,{\sc ii}]} emission in NGC 185 (from \citealt{DeLooze2016}, shown on Fig. \ref{n185-CII}) coincides with the position of the object  {VIEM} 7,  {which could cause compression of part of the shock front from  {VIEM} 8 or 9}. In addition, \cite{DeLooze2017} discuss the origin and properties of the CO and \hbox{[C\,{\sc ii}]} emission  in detail, giving density estimates of n$_{H}\sim10^4$ cm$^{-3}$, indicating the presence of a dense molecular cloud in the center of NGC 185, which is in favor of both scenarios.

\textit{ {  \bf{VIEM} 6.}} The diffuse H$\alpha$ emission of object  {VIEM} 6 \citep[PN-7 by][]{Goncalves2012} is seen in our photometric data (Fig. \ref{n185-CII}) and was only crossed once with the slit PA150 (Fig. \ref{pa150}). This low-resolution spectrum is of low quality with a low S/N (Fig. \ref{lowresspectra}, top panel), but it reveals a strong \hbox{[S\,{\sc ii}]} compared to the H$\alpha$ line. The spectrum exhibits \hbox{[S\,{\sc ii}]}/H$\alpha$ ratios of $1$, and \hbox{[N\,{\sc ii}]}/H$\alpha$ in the range $0.7 - 2$ (Fig. \ref{pa150},  {panel 4}), indicating that VIEM 6 may also be a shock-heated object, such as an SNR. {Previously catalogued as a PN by \citet{Goncalves2012}, VIEM 6 occupies shock-heated regions on all diagnostic diagrams. We note that since we barely detected the H$\beta$ line in the low-S/N spectrum of VIEM 6 (see Fig. \ref{lowresspectra}), H$\beta$ intensity represents an upper limit, and therefore the inferred  \hbox{[O\,{\sc iii}]}/H$\beta=2.5$ is a lower limit. For this object, \citet{Goncalves2012} obtained \hbox{[O\,{\sc iii}]}/H$\beta=26$, which is in significant discrepancy with our measurement and the spectra shown, but that value would place  {VIEM} 6 in the PN parameter space on BPT plots (Fig. \ref{bpt}). However, \hbox{[N\,{\sc ii}]}/H$\alpha$ and \hbox{[S\,{\sc ii}]}/H$\alpha$ ratios support the idea that  {VIEM} 6 is shock-heated (Fig. \ref{diag}).}

Finally, we note that we detected a faint diffuse ionized gas (DIG) emission  with emission line ratios of \hbox{[S\,{\sc ii}]}/H$\alpha$=1.2 and \hbox{[N\,{\sc ii}]}/H$\alpha$=0.5, which is seen far outside the central region (from $\sim$50\arcsec   to the north, to  $\sim$100\arcsec \ to the south) along the slit PA110.

\citet{DeLooze2017} made an inventory of NGC 185 gas based on \hbox{H\,{\sc i}}, CO, \textit{Spitzer}  spectra, \textit{Herschel} \hbox{[C\,{\sc ii}]}, H$\alpha,$ and X-ray observations. These latter authors concluded that NGC 185 has a lower gas-to-dust ratio than expected for low metal abundances, which requires efficient removal of a large gas fraction and a longer dust survival time. Additionally, \citet{DeLooze2017} suggested several times that shock excitation might be able to account for the detection of several tracers with high excitation temperatures; for example, shock excitation might be able to account for the observed \hbox{[O\,{\sc i}]} 63 $\mu$m emission in NGC 185 for shock velocities $\geq$ 35 km s$^{-1}$. The same model could provide at most 10 \% of the observed \hbox{[C\,{\sc ii}]} emission in NGC 185. This goes along with our assumptions that in this atypical dwarf galaxy we detected optical emission from two SNRs. 

A SN rate of $4.6\times 10^{-6}$ yr$^{-1}$ in NGC 185 was calculated based on the SFR derived by \citet{MartinezDelgado1999} (SFR=$6.6 \times 10^{-4}$ M$_{\sun}$ yr$^{-1}$), which results in a time interval of $2.2 \times 10^5$ years between SN explosions. Another assessment of SN rate, which was an
order of magnitude higher, was made by \citet{Martins2012} using chemical evolution models. This would mean that the expected interval between SN explosions would be  $1.1 \times 10^4$ years. With this higher SN rate, the existence of more than one SNR in this galaxy is possible.

\section{Summary}

In this work, we aimed to confirm the status of emission-line nebulae in the galaxy NGC 185, with special attention being devoted to SNRs. To this end, we performed photometric observations using H$\alpha$, and for the first time using the \hbox{[S\,{\sc ii}]} narrowband filter to trace shock-heated gas. Subsequently, we made follow-up spectroscopic observations with five slit positions across the central part of the galaxy,  both in low-(FWHM$\sim$500 km s$^{-1}$) and high-(FWHM$\sim$120 km s$^{-1}$)resolution modes, to infer the kinematics of the emitted gas.  We also checked archival X-ray and radio observations to search for counterparts of our optical SNRs. Our results can be summarized as follows:
\begin{itemize}
  \item[(i)] Photometric observations through  H$\alpha$ and \hbox{[S\,{\sc ii}]} narrowband filter detected 11 objects -- six previously known PNe and one PN with a possible additional source of shock ionization; one previously known symbiotic star;  {one previously known SNR, one additional optical SNR candidate; and one composite object (photoionization with some signatures of shock),   {previously cataloged as being part of an SNR}}.  
  \item[(ii)] Spectroscopic observations of SNR candidates confirmed the presence of shock-heated emission lines, which was supported by diagnostic diagrams. These confirmed the detection of two optical SNRs (VIEM 8, 9).  {The question of object  {VIEM} 7 remains open, with the most probable scenario being that it is a compact \hbox{H\,{\sc ii}} region with signatures of shock ionization}. In addition, one previously classified PN (VIEM 6) could possibly be another optical SNR candidate, but this should be taken with caution due to the poor quality of the obtained spectrum.
  \item[(iii)] High-resolution spectra showed complex kinematics of the extended emission with filaments of high expansion velocities ($\sim$50 km s$^{-1}$ 
). We were only able to observe the receding side of the expanding shell of objects  {VIEM} 8 and  {VIEM} 9, possibly due to the higher gas density on the far side of the shells.  {The absence of gas motion in object  {VIEM} 7 supports its classification as an \hbox{H\,{\sc ii}} region.}
  \item[(iv)] The estimated electron density of emission nebulae is 30 -- 200 cm$^{-3}$, which is somewhat higher than expected in the dwarf spheroidal galaxies, but is in agreement with the IR observations.
  \item[(v)] Archival {\it XMM-Newton} observations indicate the presence of an extended source in projection of our SNR candidate labeled as  {VIEM} 8. One possible  spectral model indicates 
  $0.18\ \mathrm{keV}$ CIE plasma with high intrinsic hydrogen column density, 
or  $0.27\ \mathrm{keV}$ CIE plasma with an additional power-law component 
and negligible intrinsic X-ray absorption, which may imply the presence of an unseen compact object.  {Archival VLA radio data give an indication of  weak and unresolved diffuse radio continuum emission in the center of NGC 185.}
\end{itemize}

The center of the dwarf spheroidal galaxy NGC 185 indeed hides complex nebular emission, and displays unusual emission and stellar activity for this type of galaxy. Further high-quality, multiwavelength  observations, especially in the radio and X-rays, could solidify our findings.

\begin{acknowledgements} This research has been supported by the Ministry of Education, Science  and Technological Development of the Republic of Serbia through project No. 176005 "Emission nebulae: structure and evolution" and it is a part of the joint project of Serbian Academy of Sciences and Arts and Bulgarian Academy of Sciences "Optical search for supernova remnants and H II regions in nearby galaxies". OE and AM acknowledge the support by the Russian Foundation for Basic Research grant \# 18-02-00976. OE acknowledges the support from the Program of development of M.V. Lomonosov Moscow State University (Leading Scientific School ``Physics of stars, relativistic objects and galaxies''), and from the Foundation of development of theoretical physics and mathematics ``Basis''. This study is based on observations conducted with the 6m telescope of the Special Astrophysical Observatory of the Russian Academy of Sciences carried out with the financial support of the Ministry of Science and Higher Education of the Russian Federation. 
This research has made use of the NASA/IPAC Extragalactic Database (NED) which is operated by the Jet Propulsion Laboratory, California Institute of Technology, under contract with the National Aeronautics and Space Administration, and NASA's Astrophysics Data System Bibliographic Services. This research has made use of data obtained from the 3XMM \textit{XMM-Newton} serendipitous source catalog compiled by the ten institutes of the \textit{XMM-Newton} Survey Science Centre selected by ESA, as well as the VLA archive of the National Radio Astronomy Observatory, which is a facility of the National Science Foundation operated under cooperative agreement by Associated Universities, Inc. This work is based in part on observations made with the Spitzer Space Telescope, which is operated by the Jet Propulsion Laboratory, California Institute of Technology under a contract with NASA. We thank Ilse De Looze for providing us with the \textit{Herschel} PACS \hbox{[C\,{\sc ii}]} maps.
\end{acknowledgements}

\bibliography{NGC185-final-arxiv.bbl}

\begin{thebibliography}{57}
\expandafter\ifx\csname natexlab\endcsname\relax\def\natexlab#1{#1}\fi

\bibitem[{{Afanasiev} \& {Moiseev}(2005)}]{scorpio}
{Afanasiev}, V.~L. \& {Moiseev}, A.~V. 2005, Astronomy Letters, 31, 194

\bibitem[{{Allen} {et~al.}(2008){Allen}, {Groves}, {Dopita}, {Sutherland}, \&
  {Kewley}}]{Allen2008}
{Allen}, M.~G., {Groves}, B.~A., {Dopita}, M.~A., {Sutherland}, R.~S., \&
  {Kewley}, L.~J. 2008, The Astrophysical Journal Supplement Series, 178, 20

\bibitem[{{Baade}(1951)}]{Baade1951}
{Baade}, W. 1951, Publications of Michigan Observatory, 10, 7

\bibitem[{{Baldwin} {et~al.}(1981){Baldwin}, {Phillips}, \&
  {Terlevich}}]{Baldwin1981}
{Baldwin}, J.~A., {Phillips}, M.~M., \& {Terlevich}, R. 1981, Publications of
  the Astronomical Society of the Pacific, 93, 5

\bibitem[{{Bandiera} \& {Petruk}(2004)}]{Bandiera2004}
{Bandiera}, R. \& {Petruk}, O. 2004, \aap, 419, 419

\bibitem[{{Blair} \& {Long}(1997)}]{BlairLong1997}
{Blair}, W.~P. \& {Long}, K.~S. 1997, The Astrophysical Journal Supplement
  Series, 108, 261

\bibitem[{{Bozzetto} {et~al.}(2017){Bozzetto}, {Filipovi{\'c}}, {Vukoti{\'c}},
  {Pavlovi{\'c}}, {Uro{\v{s}}evi{\'c}}, {Kavanagh}, {Arbutina}, {Maggi},
  {Sasaki}, {Haberl}, {Crawford}, {Roper}, {Grieve}, \&
  {Points}}]{Bozzetto2017}
{Bozzetto}, L.~M., {Filipovi{\'c}}, M.~D., {Vukoti{\'c}}, B., {et~al.} 2017,
  The Astrophysical Journal Supplement Series, 230, 2

\bibitem[{{Brandt} {et~al.}(1997){Brandt}, {Ward}, {Fabian}, \&
  {Hodge}}]{Brandt1997}
{Brandt}, W.~N., {Ward}, M.~J., {Fabian}, A.~C., \& {Hodge}, P.~W. 1997,
  \mnras, 291, 709

\bibitem[{{Cardelli} {et~al.}(1989){Cardelli}, {Clayton}, \&
  {Mathis}}]{Cardelli1989}
{Cardelli}, J.~A., {Clayton}, G.~C., \& {Mathis}, J.~S. 1989, \apj, 345, 245

\bibitem[{{Corradi} {et~al.}(2005){Corradi}, {Magrini}, {Greimel}, {Irwin},
  {Leisy}, {Lennon}, {Mampaso}, {Perinotto}, {Pollacco}, {Walsh}, {Walton}, \&
  {Zijlstra}}]{Corradi2005}
{Corradi}, R.~L.~M., {Magrini}, L., {Greimel}, R., {et~al.} 2005, \aap, 431,
  555

\bibitem[{{Cox}(1972)}]{Cox1972}
{Cox}, D.~P. 1972, \apj, 178, 159

\bibitem[{{Crnojevi{\'c}} {et~al.}(2014){Crnojevi{\'c}}, {Ferguson}, {Irwin},
  {McConnachie}, {Bernard}, {Fardal}, {Ibata}, {Lewis}, {Martin}, {Navarro},
  {No{\"e}l}, \& {Pasetto}}]{Crnojevic2014}
{Crnojevi{\'c}}, D., {Ferguson}, A.~M.~N., {Irwin}, M.~J., {et~al.} 2014,
  \mnras, 445, 3862

\bibitem[{{De Looze} {et~al.}(2016){De Looze}, {Baes}, {Bendo}, {Fritz},
  {Boquien}, {Cormier}, {Gentile}, {Kennicutt}, {Madden}, {Smith}, \&
  {Young}}]{DeLooze2016}
{De Looze}, I., {Baes}, M., {Bendo}, G.~J., {et~al.} 2016, \mnras, 459, 3900

\bibitem[{{De Looze} {et~al.}(2017){De Looze}, {Baes}, {Cormier}, {Kaneko},
  {Kuno}, {Young}, {Bendo}, {Boquien}, {Fritz}, {Gentile}, {Kennicutt},
  {Madden}, {Smith}, \& {Wilson}}]{DeLooze2017}
{De Looze}, I., {Baes}, M., {Cormier}, D., {et~al.} 2017, \mnras, 465, 3741

\bibitem[{{Dickel} {et~al.}(1985){Dickel}, {D'Odorico}, \&
  {Silverman}}]{Dickel1985}
{Dickel}, J.~R., {D'Odorico}, S., \& {Silverman}, A. 1985, \aj, 90, 414

\bibitem[{{Dopita} {et~al.}(1984){Dopita}, {Binette}, \& {Tuohy}}]{Dopita1984}
{Dopita}, M.~A., {Binette}, L., \& {Tuohy}, I.~R. 1984, \apj, 282, 142

\bibitem[{{Ducci} {et~al.}(2014){Ducci}, {Kavanagh}, {Sasaki}, \&
  {Koribalski}}]{ducci14}
{Ducci}, L., {Kavanagh}, P.~J., {Sasaki}, M., \& {Koribalski}, B.~S. 2014,
  \aap, 566, A115

\bibitem[{{Filipovic} {et~al.}(1998){Filipovic}, {Haynes}, {White}, \&
  {Jones}}]{Filipovic1998}
{Filipovic}, M.~D., {Haynes}, R.~F., {White}, G.~L., \& {Jones}, P.~A. 1998,
  Astronomy and Astrophysics Supplement Series, 130, 421

\bibitem[{{Ford} {et~al.}(1977){Ford}, {Jacoby}, \& {Jenner}}]{Ford1977}
{Ford}, H.~C., {Jacoby}, G., \& {Jenner}, D.~C. 1977, \apj, 213, 18

\bibitem[{{Gallagher} {et~al.}(1984){Gallagher}, {Hunter}, \&
  {Mould}}]{Gallagher1984}
{Gallagher}, J.~S., I., {Hunter}, D.~A., \& {Mould}, J. 1984, \apj, 281, L63

\bibitem[{{Ge} {et~al.}(2015){Ge}, {Li}, {Xu}, {Gu}, {Wang}, {Roberts},
  {Kraft}, {Jones}, \& {Forman}}]{Ge2015}
{Ge}, C., {Li}, Z., {Xu}, X., {et~al.} 2015, \apj, 812, 130

\bibitem[{{Geha} {et~al.}(2015){Geha}, {Weisz}, {Grocholski}, {Dolphin}, {van
  der Marel}, \& {Guhathakurta}}]{Geha2015}
{Geha}, M., {Weisz}, D., {Grocholski}, A., {et~al.} 2015, \apj, 811, 114

\bibitem[{{Gon{\c{c}}alves} {et~al.}(2012){Gon{\c{c}}alves}, {Magrini},
  {Martins}, {Teodorescu}, \& {Quireza}}]{Goncalves2012}
{Gon{\c{c}}alves}, D.~R., {Magrini}, L., {Martins}, L.~P., {Teodorescu}, A.~M.,
  \& {Quireza}, C. 2012, \mnras, 419, 854

\bibitem[{{Heckman} {et~al.}(1980){Heckman}, {Balick}, \&
  {Crane}}]{Heckman1980}
{Heckman}, T.~M., {Balick}, B., \& {Crane}, P.~C. 1980, Astronomy and
  Astrophysics Supplement Series, 40, 295

\bibitem[{{Ho} {et~al.}(1997){Ho}, {Filippenko}, \& {Sargent}}]{Ho1997}
{Ho}, L.~C., {Filippenko}, A.~V., \& {Sargent}, W. L.~W. 1997, The
  Astrophysical Journal Supplement Series, 112, 315

\bibitem[{{Ho} \& {Ulvestad}(2001)}]{HoUlvestad2001}
{Ho}, L.~C. \& {Ulvestad}, J.~S. 2001, The Astrophysical Journal Supplement
  Series, 133, 77

\bibitem[{{Kauffmann} {et~al.}(2003){Kauffmann}, {Heckman}, {Tremonti},
  {Brinchmann}, {Charlot}, {White}, {Ridgway}, {Brinkmann}, {Fukugita}, {Hall},
  {Ivezi{\'c}}, {Richards}, \& {Schneider}}]{Kauffmann2003}
{Kauffmann}, G., {Heckman}, T.~M., {Tremonti}, C., {et~al.} 2003, \mnras, 346,
  1055

\bibitem[{{Kewley} {et~al.}(2001){Kewley}, {Dopita}, {Sutherland}, {Heisler},
  \& {Trevena}}]{Kewley2001}
{Kewley}, L.~J., {Dopita}, M.~A., {Sutherland}, R.~S., {Heisler}, C.~A., \&
  {Trevena}, J. 2001, \apj, 556, 121

\bibitem[{{Long}(2017)}]{Long2017}
{Long}, K.~S. 2017, {Galactic and Extragalactic Samples of Supernova Remnants:
  How They Are Identified and What They Tell Us}, 2005

\bibitem[{{Markwardt}(2009)}]{mpfit}
{Markwardt}, C.~B. 2009, in Astronomical Society of the Pacific Conference
  Series, Vol. 411, Astronomical Data Analysis Software and Systems XVIII, ed.
  D.~A. {Bohlender}, D.~{Durand}, \& P.~{Dowler}, 251

\bibitem[{{Marleau} {et~al.}(2010){Marleau}, {Noriega-Crespo}, \&
  {Misselt}}]{Marleau2010}
{Marleau}, F.~R., {Noriega-Crespo}, A., \& {Misselt}, K.~A. 2010, \apj, 713,
  992

\bibitem[{{Mart{\'\i}nez-Delgado} {et~al.}(1999){Mart{\'\i}nez-Delgado},
  {Aparicio}, \& {Gallart}}]{MartinezDelgado1999}
{Mart{\'\i}nez-Delgado}, D., {Aparicio}, A., \& {Gallart}, C. 1999, \aj, 118,
  2229

\bibitem[{{Martins} {et~al.}(2012){Martins}, {Lanfranchi}, {Gon{\c{c}}alves},
  {Magrini}, {Teodorescu}, \& {Quireza}}]{Martins2012}
{Martins}, L.~P., {Lanfranchi}, G., {Gon{\c{c}}alves}, D.~R., {et~al.} 2012,
  \mnras, 419, 3159

\bibitem[{{Mateo}(1998)}]{1998Mateo}
{Mateo}, M.~L. 1998, \araa, 36, 435

\bibitem[{{Matonick} \& {Fesen}(1997)}]{MatonickFesen1997}
{Matonick}, D.~M. \& {Fesen}, R.~A. 1997, The Astrophysical Journal Supplement
  Series, 112, 49

\bibitem[{{McConnachie}(2012)}]{McConnachie2012}
{McConnachie}, A.~W. 2012, \aj, 144, 4

\bibitem[{{Monet}(1998)}]{Monet1998}
{Monet}, D. 1998, {USNO-A2.0}

\bibitem[{{Oke}(1990)}]{Oke1990}
{Oke}, J.~B. 1990, \aj, 99, 1621

\bibitem[{{Osterbrock}(1989)}]{Osterbrock}
{Osterbrock}, D.~E. 1989, {Astrophysics of gaseous nebulae and active galactic
  nuclei}

\bibitem[{{Richer} \& {McCall}(2008)}]{RicherMcCall2008}
{Richer}, M.~G. \& {McCall}, M.~L. 2008, \apj, 684, 1190

\bibitem[{{Rosen} {et~al.}(2016){Rosen}, {Webb}, {Watson}, {Ballet}, {Barret},
  {Braito}, {Carrera}, {Ceballos}, {Coriat}, {Della Ceca}, {Denkinson},
  {Esquej}, {Farrell}, {Freyberg}, {Gris{\'e}}, {Guillout}, {Heil},
  {Koliopanos}, {Law-Green}, {Lamer}, {Lin}, {Martino}, {Michel}, {Motch},
  {Nebot Gomez-Moran}, {Page}, {Page}, {Page}, {Pakull}, {Pye}, {Read},
  {Rodriguez}, {Sakano}, {Saxton}, {Schwope}, {Scott}, {Sturm}, {Traulsen},
  {Yershov}, \& {Zolotukhin}}]{3XMM}
{Rosen}, S.~R., {Webb}, N.~A., {Watson}, M.~G., {et~al.} 2016, \aap, 590, A1

\bibitem[{{Sasaki} {et~al.}(2018){Sasaki}, {Haberl}, {Henze}, {Saeedi},
  {Williams}, {Plucinsky}, {Hatzidimitriou}, {Karampelas}, {Sokolovsky},
  {Breitschwerdt}, {de Avillez}, {Filipovi{\'c}}, {Galvin}, {Kavanagh}, \&
  {Long}}]{SHH18}
{Sasaki}, M., {Haberl}, F., {Henze}, M., {et~al.} 2018, \aap, 620, A28

\bibitem[{{Schlafly} \& {Finkbeiner}(2011)}]{Schlafly2011}
{Schlafly}, E.~F. \& {Finkbeiner}, D.~P. 2011, \apj, 737, 103

\bibitem[{{Sedov}(1959)}]{Sedov1959}
{Sedov}, L.~I. 1959, {Similarity and Dimensional Methods in Mechanics}

\bibitem[{{Smith} {et~al.}(2001){Smith}, {Brickhouse}, {Liedahl}, \&
  {Raymond}}]{apec01}
{Smith}, R.~K., {Brickhouse}, N.~S., {Liedahl}, D.~A., \& {Raymond}, J.~C.
  2001, \apjl, 556, L91

\bibitem[{{Snowden} \& {Kuntz}(2011)}]{SK11}
{Snowden}, S.~L. \& {Kuntz}, K.~D. 2011, in Bulletin of the American
  Astronomical Society, Vol.~43, American Astronomical Society Meeting
  Abstracts \#217, 344.17

\bibitem[{{Stasi{\'n}ska} {et~al.}(2006){Stasi{\'n}ska}, {Cid Fernandes},
  {Mateus}, {Sodr{\'e}}, \& {Asari}}]{Stasinska2006}
{Stasi{\'n}ska}, G., {Cid Fernandes}, R., {Mateus}, A., {Sodr{\'e}}, L., \&
  {Asari}, N.~V. 2006, \mnras, 371, 972

\bibitem[{{Sturm} {et~al.}(2013){Sturm}, {Haberl}, {Pietsch}, {Ballet},
  {Hatzidimitriou}, {Buckley}, {Coe}, {Ehle}, {Filipovi{\'c}}, {La Palombara},
  \& {Tiengo}}]{SHP13}
{Sturm}, R., {Haberl}, F., {Pietsch}, W., {et~al.} 2013, \aap, 558, A3

\bibitem[{{Taylor}(1950)}]{Taylor1950}
{Taylor}, G. 1950, Proceedings of the Royal Society of London Series A, 201,
  159

\bibitem[{{Tolstoy} {et~al.}(2003){Tolstoy}, {Venn}, {Shetrone}, {Primas},
  {Hill}, {Kaufer}, \& {Szeifert}}]{2003TolstoyVenn}
{Tolstoy}, E., {Venn}, K.~A., {Shetrone}, M., {et~al.} 2003, \aj, 125, 707

\bibitem[{{Vargas} {et~al.}(2014){Vargas}, {Geha}, \& {Tollerud}}]{Vargas2014}
{Vargas}, L.~C., {Geha}, M.~C., \& {Tollerud}, E.~J. 2014, \apj, 790, 73

\bibitem[{{Vu{\v{c}}eti{\'c}} {et~al.}(2016){Vu{\v{c}}eti{\'c}}, {Arbutina},
  {Pavlovic}, {Ciprijanovic}, {Urosevic}, {Petrov}, {Oni{\'c}}, \&
  {Trcka}}]{Vucetic2016}
{Vu{\v{c}}eti{\'c}}, M., {Arbutina}, B., {Pavlovic}, M.~Z., {et~al.} 2016, in
  Supernova Remnants: An Odyssey in Space after Stellar Death, 34

\bibitem[{{Vu{\v{c}}eti{\'c}} {et~al.}(2015){Vu{\v{c}}eti{\'c}}, {Arbutina}, \&
  {Uro{\v{s}}evi{\'c}}}]{Vucetic2015}
{Vu{\v{c}}eti{\'c}}, M.~M., {Arbutina}, B., \& {Uro{\v{s}}evi{\'c}}, D. 2015,
  \mnras, 446, 943

\bibitem[{{Vu{\v{c}}eti{\'c}} {et~al.}(2013){Vu{\v{c}}eti{\'c}}, {Arbutina},
  {Uro{\v{s}}evi{\'c}}, {Dobardzic}, {Pavlovic}, {Pannuti}, \&
  {Petrov}}]{Vucetic2013}
{Vu{\v{c}}eti{\'c}}, M.~M., {Arbutina}, B., {Uro{\v{s}}evi{\'c}}, D., {et~al.}
  2013, Serbian Astronomical Journal, 187, 11

\bibitem[{{White} {et~al.}(2019){White}, {Long}, {Becker}, {Blair}, {Helfand},
  \& {Winkler}}]{White2019}
{White}, R.~L., {Long}, K.~S., {Becker}, R.~H., {et~al.} 2019, arXiv e-prints,
  arXiv:1903.04434

\bibitem[{{Wilms} {et~al.}(2000){Wilms}, {Allen}, \& {McCray}}]{tbabs00}
{Wilms}, J., {Allen}, A., \& {McCray}, R. 2000, \apj, 542, 914

\bibitem[{{Young} \& {Lo}(1997)}]{YoungLo1997}
{Young}, L.~M. \& {Lo}, K.~Y. 1997, \apj, 476, 127

\end{thebibliography}


\end{document}